\documentclass[reqno,12pt]{article}
\usepackage{amsthm,amsmath,amssymb,graphicx,setspace,verbatim,natbib}
\usepackage[small]{caption}
\usepackage{subcaption}
\usepackage{wrapfig}
\usepackage{multirow}
\usepackage{rotating}
\usepackage{color}
\usepackage{changepage}

\renewcommand{\arraystretch}{.95}

\newcommand{\bzero}{\mbox{\boldmath $0$}}

\newcommand{\bb}{\mbox{\boldmath $b$}}

\newcommand{\bv}{\mbox{\boldmath $v$}}

\newcommand{\bw}{\mbox{\boldmath $w$}}
\newcommand{\bx}{\mbox{\boldmath $x$}}
\newcommand{\tbx}{\tilde{\mbox{\boldmath $x$}}}
\newcommand{\tby}{\tilde{\mbox{\boldmath $y$}}}
\newcommand{\by}{\mbox{\boldmath $y$}}
\newcommand{\bz}{\mbox{\boldmath $z$}}

\newcommand{\bI}{\mbox{\boldmath $I$}}

\newcommand{\bQ}{\mbox{\boldmath $Q$}}

\newcommand{\bV}{\mbox{\boldmath $V$}}

\newcommand{\bX}{\mbox{\boldmath $X$}}

\newcommand{\cG}{{\cal G}}

\newcommand{\cI}{{\cal I}}

\newcommand{\cM}{{\cal M}}
\newcommand{\cN}{{\cal N}}

\newcommand{\cW}{{\cal W}}

\newcommand{\new}{\mbox{\scriptsize{new}}}

\newcommand{\ty}{\tilde{y}}

\newcommand{\tq}{\tilde{q}}

\newcommand{\hs}{\hat{s}}

\newcommand{\hlambda}{\hat{\lambda}}

\newcommand{\tsigma}{\tilde{\sigma}}

\newcommand{\bbeta}{\mbox{\boldmath $\beta$}}

\newcommand{\btheta}{\mbox{\boldmath $\theta$}}

\newcommand{\bsx}{\mbox{\scriptsize \boldmath $x$}}

\newcommand{\bphi}{\mbox{\boldmath $\phi$}}

\newcommand{\bgamma}{\mbox{\boldmath $\gamma$}}

\newcommand{\bmu}{\mbox{\boldmath $\mu$}}
\newcommand{\tbmu}{\tilde{\mbox{\boldmath $\mu$}}}
\newcommand{\bpi}{\mbox{\boldmath $\pi$}}

\newcommand{\bSigma}{\mbox{\boldmath $\Sigma$}}
\newcommand{\tbSigma}{\tilde{\mbox{\boldmath $\Sigma$}}}
\newcommand{\tbD}{\tilde{\mbox{\boldmath $D$}}}
\newcommand{\tbQ}{\tilde{\mbox{\boldmath $Q$}}}

\newcommand{\bPsi}{\mbox{\boldmath $\Psi$}}

\newcommand{\bOmega}{\mbox{\boldmath $\Omega$}}

\newcommand{\bnu}{\mbox{\boldmath $\nu$}}

\newcommand{\tbb}{\tilde{\mbox{\boldmath $b$}}}

\newcommand{\E}{\mbox{E}}
\newcommand{\Var}{\mbox{Var}}

\newcommand{\miss}{\mbox{\scriptsize miss}}

\newcommand{\trigamma}{\digamma'}
\newcommand{\pr}{\mbox{\scriptsize p}}

\newcommand{\bdm}{\begin{displaymath}}
\newcommand{\edm}{\end{displaymath}}
\newcommand{\beq}{\begin{equation}}
\newcommand{\eeq}{\end{equation}}

\renewcommand{\th}{^{\mbox{\scriptsize th}}}

\long\def\symbolfootnote[#1]#2{\begingroup%
\def\thefootnote{\fnsymbol{footnote}}\footnote[#1]{#2}\endgroup}

\newlength{\offsetpage}
\newenvironment{widepage}{\begin{adjustwidth}{-\offsetpage}{-\offsetpage}%
    \addtolength{\textwidth}{2\offsetpage}}%
{\end{adjustwidth}}

\begin{document}

\pagenumbering{arabic}
\begin{center}
{\singlespacing
\begin{Large}{\bf
Prediction and Inference with Missing Data in Patient Alert Systems\\}
\vspace{.25in}
\end{Large}

Curtis$\:$B.$\:$Storlie, Terry$\:$M.$\:$Therneau, Rickey$\:$E.$\:$Carter, Nicholas Chia, John$\:$R.$\:$Bergquist, \\Jeanne$\:$M.$\:$Huddleston, Santiago Romero-Brufau\\[.25in]

Mayo Clinic\\[.3in]

\begin{abstract}

We describe the Bedside Patient Rescue (BPR) project, the goal of which is risk prediction of adverse events for non-ICU patients using $\sim$200 variables (vitals, lab results, assessments, ...).  There are several missing predictor values for most patients, which in the health sciences is the norm, rather than the exception.  A Bayesian approach is presented that addresses many of the shortcomings to standard approaches to missing predictors: (i) treatment of the uncertainty due to imputation is straight-forward in the Bayesian paradigm, (ii) the predictor distribution is flexibly modeled as an infinite normal mixture with latent variables to explicitly account for discrete predictors (i.e., as in multivariate probit regression models), and (iii) certain missing not at random situations can be handled effectively by allowing the indicator of missingness into the predictor distribution only to inform the distribution of the missing variables. The proposed approach also has the benefit of providing a distribution for the prediction, including the uncertainty inherent in the imputation. Therefore, we can ask questions such as: is it possible this individual is at high risk but we are missing too much information to know for sure?  How much would we reduce the uncertainty in our risk prediction by obtaining a particular missing value?  This approach is applied to the BPR problem resulting in excellent predictive capability to identify deteriorating patients.

\vspace{.06in}

\vspace{.06in}
\noindent
{\em Running title}: Prediction and Inference with Missing Data in Patient Alert Systems

\vspace{.06in}
\noindent
{\em Corresponding Author}: Curtis Storlie, \verb1storlie.curt@mayo.edu1

\end{abstract}
}

\end{center}


\vspace{-.35in}
\section{Introduction}
\vspace{-.09in}

The Bedside Patient Rescue (BPR) project at Mayo Clinic is an automated alert system to predict risk of deterioration of patients in general care floors.
The primary response of interest was the time until one of the following events (i.e., cardiac arrest, transfer to the intensive care unit (ICU), or the patient requiring rapid response team intervention).  
Hospital patients typically show signs of deterioration up to days prior to adverse outcomes like cardiorespiratory arrest \citep{Buist1999,Schein1990}. The rate of in-hospital cardiorespiratory arrest (CRA) requiring cardiopulmonary resuscitation is estimated to be 0.174 per bed per year in the US \citep{Peberdy2003}. After a cardiac arrest, survival to discharge is estimated to be as low as 18\% \citep{Peberdy2003,Nadkarni2006}, therefore, efforts to predict and prevent arrest could prove beneficial \citep{Buist1999,Schein1990}.  There have been several proposed approaches for early warning systems \citep{Kirkland2013,Griffiths2011,Paterson2006,Duckitt2007,Prytherch2010,Smith2013}, however, existing risk scores (called)automated approaches have been hindered by low positive predictive values \citep{Romero2014}.

The BPR project is unique in that it uses many ($\sim$200) predictors (vital signs, lab results, nursing assessments, demographics,$\dots$) to create a risk score.  The training data was extracted from patients at Mayo Clinic from 2010$-$2011. Data from each patient with an event was extracted.  For each event, data from ten other randomly selected patients in the hospital the same day as that event were also extracted.  Many of the variables are time varying, and some are time-lagged vitals (e.g., min and max in the last 24 hours).  However, due to engineering decisions made at the time of implementation, the entire time profiles for vitals, etc., are not available for the patients.  Several of the candidate predictor variables are the result of clinician selected multiplicative interactions and ratios (e.g., the shock index is heart rate over blood pressure). The predictor variables used are described in more detail in Section~\ref{sec:BPR} and the complete list of 171 variables collected and calculated on each patient are provided in the supplementary material.

The focus of this paper is on the statistical modeling and analysis of the BPR data, and in particular the treatment of missing data.  Missing data problems are very common in practice, particularly in health sciences. All patients in the BPR training data set have a missing value for at least one of their predictors.  The average number of missing values for patients is 25, many of these variables being lab tests such as albumin or troponin, which are informative risk factors.  Thus, a simple approach of excluding cases or variables that have missing data is not practical.  Another common approach is to create a “missing” indicator and include it in the regression.  Similarly, tree based algorithms like Gradient Boosting Machine (GBM) \citep{Friedman01} treat missing values as a separate (third) node in a split on a variable and work well for prediction.
However, with the above approaches, interpretation becomes challenging and this does not leverage the relationships among predictors.  Regression based imputation, e.g., Random Forest Imputation \citep{Stekhoven2012}, can result in good prediction in many cases, however, it assumes no uncertainty about the imputed values making it difficult to assess uncertainty in predictions and inference.

\begin{figure}[t!]
\vspace{-.13in}
\centering
\caption{Pairwise scatter plots of a few BPR predictors (a random sample of 2000 observations).}
\vspace{-.15in}
\includegraphics[width=.93\textwidth]{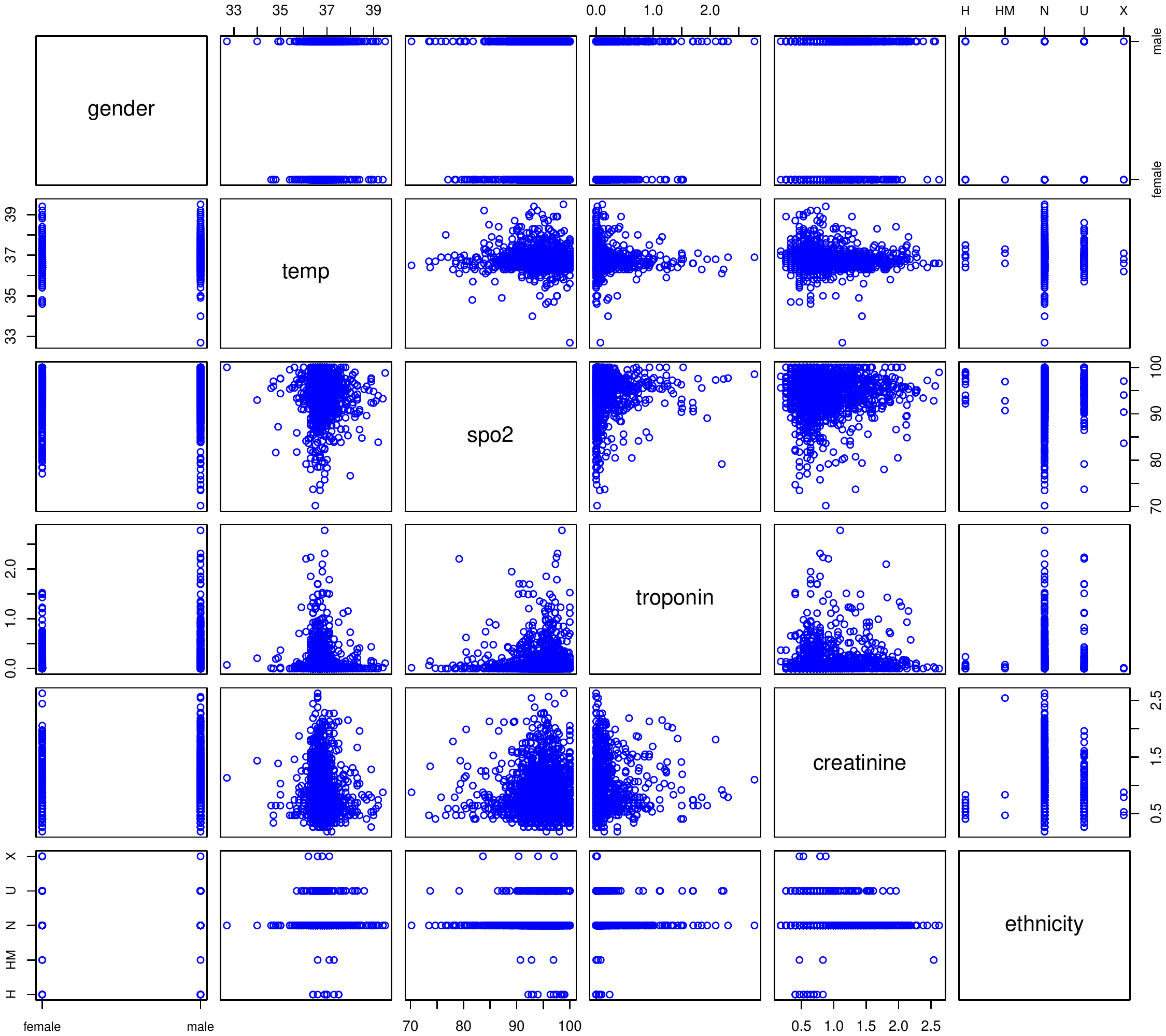}
\label{fig:pw_BPR}
\vspace{-.1in}
\end{figure}

Multiple imputation (MI) pioneered by \cite{Rubin1976} is an intuitively attractive concept to admit this uncertainty along with maximum likelihood (ML) of the observed data \citep[e.g.,][]{Schafer1997}.  MI and ML methods typically make the commonly misunderstood assumption of missing at random (MAR) \cite{Rubin1976}.  The MAR assumption implies that the likelihood of a missing value {\em can} depend on the value of the unobserved variable, marginally, just not after accounting for all observed variables.
Both MI and ML essentially aim to marginalize over the distribution of the missing data.  The main caveat is that care must be taken to accurately represent the joint distribution of the {\em complete} data, i.e., that for all variables in the data set.  The prominent approach is to assume a multivariate normal (MVN) distribution for the predictors and the assumed regression model for the response(s) conditional on them.  However, predictors are commonly not Gaussian; often many are not even continuous and or have hard limits/boundaries.  In particular, this is the case for the BPR problem as seen in the pairwise scatterplots of a few of the predictors presented in Figure~\ref{fig:pw_BPR}.
Still, it is commonplace to treat categorical data as continuous, i.e., MVN, and impute them anyhow.  Such methods can work well for certain problems but are known to fail in others \citep{Allison2000,Horton2003,Bernaards2007,Finch2010,Yucel2011,Galati2014}.

An alternative to the MVN assumption and its shortcomings is to specify a sequence of univariate models for each variable conditional on all other variables.  This is known as the chained equations or fully-conditional approach \citep{Raghunathan2001,vanBuuren2006,vanBuuren2007}.  This approach is much more flexible for discrete variables, e.g., a categorical variable could be represented by a multinomial logistic or probit regression model on the other variables.  The values are then imputed one at a time in an iterative fashion resembling a Gibbs sampler.  This approach has shown practical success in many situations, however, the resistance to more widespread use is that the full conditional models may not be compatible.  That is, they may not correspond to {\em any} multivariate distribution \citep{Raghunathan2001}, which raises serious theoretical concerns and can lead to convergence issues \citep{Li2012,Si2013}.

Postulating a valid multivariate distribution is notoriously challenging in the presence of both continuous and categorical variables.  One solution is the conditional Gaussian approach \citep{Schafer1997}, where a log-linear model is specified for the discrete random variables, and conditional on this distribution, a multivariate normal distribution is assumed for the continuous variables.  However, even a modest number of categorical variables can lead to known difficulties in the estimation of log-linear models \citep{Horton2007,Si2013}.  Mixtures of independent multinomial distributions have previously been used for MI of categorical data to alleviate this problem \citep{Vermunt2008,Gebregziabher2010,Si2013}.  Recently \cite{Murray2015} have extended this approach to handle mixed categorical and continuous predictors with a conditional Gaussian framework, melding mixtures of independent multinomials for the categorical variables and mixtures of MVN conditional on the categorical values for the continuous variables.  However, with a large number of discrete variables, a very large number of mixture components will be needed to represent even simple dependencies between variables.

In this work, we extend the multivariate probit approach to modeling discrete variables \citep{Lesaffre1991,Lesaffre1992,Chib1998,Gibbons1998} to allow for a multivariate representation of categorical and continuous variables as suggested by \cite{Dunson2000} and \cite{Gueorguieva2001}.
%
A joint model for the predictors $\bx=[x_1,\dots,x_p]'$ is created by assuming a MVN model for $\bx^*$, the collection of continuous predictors and (continuous) latent variables for categorical predictors.
%
%
%
We then relax the MVN assumption by using a Dirichlet process model (DPM) (i.e., an infinite normal mixture) for the joint distribution for $\bx^*$.  As it is difficult to fit mixture models for high dimensional $\bx^*$ we use a variable selection approach \citep{Raftery2006,Kim2006,Storlie17b} to allow the component parameters to vary on only sparse subset of the $x_j$.  Despite the flexibility and broad utility for such an approach, to the best of our knowledge, it has not been used to model mixed categorical and continuous data.
We also investigate the inclusion of an indicator for missingness variable (as a continuous latent variable) into the model for $\bx^*$.  This essentially amounts to formulating a {\em selection model} \citep{Heckman1976,Amemiya1984}, where here the selection probabilities (i.e., probabilities of observing the variables) depend on the rest of $\bx^*$.  These indicators are {\em not} part of the regression model for $y$, rather they are only to inform the distribution of the missing variables.  Certain types of missing not at random (MNAR) situations can be handled effectively in this manner.
%
%

We use a Bayesian approach to estimate this model, which has similarities to both MI and ML approaches in that it aims to marginalize over the missing data and obtain the posterior distribution of the relevant parameters and/or variables, conditional on only the observed data.  This is typically accomplished by sampling the missing values inside of the Markov Chain Monte Carlo (MCMC) algorithm in a Gibbs type framework; conceptually each MCMC iteration contains two steps (i) complete the data conditional on the model parameters, and (ii) update the parameters conditional on complete data.
One of the advantages of Bayesian analysis to this problem is that it provides natural measures of uncertainty for parameters via the posterior distribution.  An important consequence of this to the BPR problem is that risk prediction for a given patient is represented as a distribution, including the uncertainty due to the missing predictors. Therefore, we can ask questions, such as, 
how much could we reduce the uncertainty in our risk prediction by obtaining a particular lab value? 

The rest of the paper is laid out as follows. Section~\ref{sec:model} describes the proposed Bayesian missing data approach for mixed categorical and continuous variables.  Section~\ref{sec:sims} evaluates the performance of this approach when compared to some other methods on several simulation cases.  The approach is then applied to the problem for which it was designed in Section~\ref{sec:BPR} where a comprehensive analysis of the BPR problem is presented.  Section~\ref{sec:conclusions} concludes the paper.  This paper also has online supplementary material containing details of the MCMC algorithm and descriptions of the variables in the BPR dataset.



\vspace{-.05in}
{\singlespacing
\section{A Bayesian Framework for Missing Data with Mixed Categorical and Continuous Variables}
}
\vspace{-.09in}
\label{sec:model}

\vspace{-.05in}
\subsection{Complete Data Model}
\vspace{-.1in}
\label{sec:model_descr}

Assume a regression model for the response $y$ as a function of predictors $\bx=[x_{1},\dots,x_{p}]'$,
\vspace{-.25in}\beq
y \mid \bx \sim f(\bx,\btheta),
\label{eq:reg_model}
\vspace{-.05in}\eeq
where $\btheta$ is a vector of parameters.  In the BPR problem it is assumed that
$y$ is the time to event and thus a scalar, but it need not be for the discussion to follow.  The specific regression model to be used for BPR is presented later in Section~\ref{sec:reg_model}.  Let the $i\th$ observation of $\bx$ and $y$ be denoted $\bx_i =[x_{i,1},\dots,x_{i,p}]'$, $i=1,\dots,n$, and $y_i$, respectively.  Some of the $x_{i,j}$ (and possibly $y_i$) are missing and thus a model for the joint distribution $F_{\bsx}$ of $\bx$ is postulated below.  The regression relation in (\ref{eq:reg_model}) along with $F_{\bsx}$ will specify the joint distribution function $F$ of $(\bx,y)$ and it is assumed that $(\bx_i,y_i) \stackrel{iid}{\sim} F$.
The predictors $\bx$ consist of a mix of categorical and continuous random variables.  With out loss of generality, assume they are ordered such that $x_j$ is categorical with $L_j \geq 2$ levels for $j=1,\dots,r \leq p$, and $x_j$ is continuous for $j=r+1,\dots,p$.  For ease of presentation, we assume the categorical variables are unordered, but the ordinal case can be handled with a straight-forward extension of the approach described below.

We first define a marginal distribution for a categorical $x_j$ as a multinomial probit model and then discuss how to build a joint distribution for $\bx$ with these marginals.  Following \cite{McCulloch2000}, the discrete variable $x_{j} \in \{0,\dots,L_j-1\}$ is modeled in terms of a latent vector $\bw_{j} = [w_{j,1},\dots,w_{j,L_j-1}]$ via the relation,
\vspace{-.15in}\bdm
x_{j}\left(\bw_{j}\right) = \left\{
\begin{array}{ll}
  0 & \mbox{if $\max(\bw_{j}) \leq 0$} \\
  k & \mbox{if $\max(\bw_{j}) = w_{j,k} > 0$}, \\
\end{array}
\right.
\vspace{-.15in}\edm
where
\vspace{-.07in}\beq
\bw_j \sim \mbox{N}(\bmu_j, \bSigma_{jj}),
\label{eq:w_j_cat}
\vspace{-.05in}\eeq
$\bmu_j$ is a mean vector of length $L_j-1$, and $\bSigma_{jj}$ with $(l,m)\th$ element $\sigma_{lm}$ is a covariance matrix.  It is well known that this model is not likelihood identified, so for identifiability purposes it is usually assumed that $\sigma_{11}=1$.

For now, suppose the marginal distribution of a continuous variable $x_j$ is Gaussian; this assumption is relaxed in Section~\ref{sec:DPM_model}.  Some continuous variables may have finite limits, e.g., a lower limit of 0 for a lab value, or an upper limit of 100\% for oxygen saturation as in Figure~\ref{fig:pw_BPR}.  Thus, for a continuous variable $x_j$ with lower and upper limits of $b_j$ and $c_j$ (which could be infinite) it is assumed that
\vspace{-.17in}\bdm
x_j\left(\bw_{j}\right) = \bw_j I_{\{b_j \leq \bw_j \leq c_j\}} + b_jI_{\{\bw_j < b_j\}} + c_jI_{\{\bw_j > c_j\}},
\vspace{-.12in}\edm
where $I_{\{A\}}=1$ if $A$ and 0 otherwise.  That is, if there are finite limits for $x_j$, then $x_j$ is assumed to be a left and/or right censored version of $\bw_j$, thus producing a positive mass at the boundary value(s) of $x_j$.  For continuous $x_j$, the $\bw_j$, $\bmu_j$, and $\bSigma_{jj}$ are all scalars, but they are in bold to maintain consistent notation to that in (\ref{eq:w_j_cat}).

The marginal distribution for each (latent) variable vector $\bw_j$, which defines the marginal distribution of each $x_j$, respectively, was assumed above to be MVN.  It is then straight-forward to define a valid joint distribution for $\bx^*=[\bw_1',\dots,\bw_p']'$, and thus for $\bx$, with the specified marginals as
\vspace{-.2in}\beq
\bx^* \sim \mbox{N}(\bmu, \bSigma)
\label{eq:w_distn}
\vspace{-.30in}\eeq
where $\bmu = [\bmu_1',\dots,\bmu_p']'$, and
\vspace{-.3in}\bdm
\bSigma = \left(
\begin{array}{ccc}
  \bSigma_{11}  & \!\cdots \!& \bSigma_{1p} \\[-.1in]
  \vdots & \!\ddots \!& \vdots \\[-.1in]
  \bSigma_{p1} & \!\cdots \!& \bSigma_{pp} 
\end{array}
\right).
\vspace{-.2in}\edm

\noindent
Denote the corresponding precision matrix as $\bQ=\bSigma^{-1}$.
Let the $(l,m)\th$ element of $\bSigma_{jk}$ and $\bQ_{jk}$ be denoted by $\sigma_{jklm}$ and $q_{jklm}$, respectively.  In order to maintain identifiability it is assumed here that $q_{jj11}=1$ for all $j \leq r$ (i.e., for all categorical $x_j$).  It would be more conventional to assume that $\sigma_{jj11}=1$, however, this restriction is arbitrary, and placing it on the precision leads to a more convenient computation.  The distribution on $\bx$ implied in (\ref{eq:w_distn}) is a natural extension of the multivariate probit model for binary variables in \cite{Chib1998} to allow for a multivariate representation of categorical variables \citep{Zhang08,Bhattacharya12}.

The restriction that $q_{jj11}=1$ for all $j \leq r$ complicates posterior inference as there is no longer a conjugate distribution for the restricted $\bQ$, which can be of large dimension and must be positive definite.  However, this problem has been relatively well studied in the multinomial probit and multivariate probit settings and various proposed solutions exist \citep{McCulloch1994,Nobile1998,McCulloch2000,Imai2005}.
In the spirit of the prior used by \cite{Imai2005} for a single multinomial probit regression model, we generalize to the multiple categorical variable setting by assuming prior distributions for $\bmu$ and $\bQ$ as 
\vspace{-.2in}\begin{eqnarray}
  \bQ & = & \tbD \tbQ \tbD, \label{eq:Sigma_prior} \\[-.07in]
  \tbQ & \sim & \mbox{Wishart}(\eta,\bOmega),\nonumber \\[-.07in]
  \bmu & \sim & \mbox{N}(\bnu,(\varphi\bQ)^{-1}), \label{eq:mu_prior}\\[-.58in] \nonumber
\end{eqnarray}
where  and $\tbD$ is a diagonal matrix to be defined below that scales $\tbQ$ to ensure $q_{jj11}=1$ for all $j \leq r$, while keeping the same correlation structure as that implied by $\tbQ$.  Let $\tq_{jklm}$ and $d_{jklm}$ represent the individual elements of $\tbQ$ and $\tbD$, respectively, in analogous fashion to $q_{jklm}$ for $\bQ$.  Then
\vspace{-.15in}\bdm
d_{jklm} = \left\{
\begin{array}{ll}
  0 & \mbox{if $j \neq k$ or $l \neq m$}\\[-.07in]
  \left(\tq_{jj11}\right)^{-\frac{1}{2}} & \mbox{$l=m=1$ and $j = k \leq r$} \\[-.07in]
  1 & \mbox{otherwise}.\\
\end{array}
\right.
\vspace{-.13in}\edm
Now define the transforms $\tbx^* = \tbD \bx^*$ and $\tbmu = \tbD \bmu$ and the prior distribution implied on the transformed variables $(\tbmu, \tbQ)$ is $\tbmu \mid \tbQ \sim \mbox{N}(\tbD \bnu, \tbQ)$ and $\tbQ \sim \mbox{Wishart}(\eta,\bOmega)$.  As in \cite{Imai2005}, this formulation allows for direct prior specification on the identifiable parameter $\bmu$ in (\ref{eq:mu_prior}) so that a either an informative or flat prior on $\bmu$ is possible.  The prior distribution for the identified $\bQ$ in (\ref{eq:Sigma_prior}) is specified indirectly via the unidentified $\tbQ$ and must be proper.  However, this prior is generally not meant to convey substantive information but rather to be weakly informative \citep{Imai2005}.

The MCMC sampling is described in more detail in Section~\ref{sec:MCMC}, but the general approach of the parameter expansion strategy is to work with the transformed variables $\tbx^*$, $\tbmu$, and $\tbQ$.  For a given $\tbx^*$, $\tbmu$ and $\tbQ$ are identified and have the usual normal-Wishart conjugate update.  Once they are updated, the likelihood identified parameters $\bmu$ and $\bQ$ are simply computed via the deterministic relations $\bmu = \tbD^{-1}\tbmu$ and $\bQ = \tbD\tbQ \tbD$.

\vspace{-.15in}
\subsection{Dirichlet Process Extension for the Complete Data Model}
\vspace{-.07in}
\label{sec:DPM_model}

The model for $\bx$ in Section~\ref{sec:model_descr} assumes that $\bx^* \sim \cN(\bmu, \bQ^{-1})$ which may  be a reasonable approximation in some cases, but often times data are not MVN as already highlighted in Figure~\ref{fig:pw_BPR}.  In many cases, variable transformations may help but this is not always the case, particularly if the data are multimodal.  In addition, a MVN model for $\bx^*$ implicitly assumes the conditional relationship between $\bw_j$ and the remaining $\bw_k$ is linear (e.g., no interactions).  An approach to alleviate these concerns is to allow $\bx^*$ to be a (infinite) Gaussian mixture via a Dirichlet process model (DPM).  That is, assume 
\vspace{-.1in}\beq
\bx^* \sim \sum_{h=1}^\infty \pi_h \:\cN(\bmu_h, \bQ^{-1}_h), 
\label{eq:DPM_model}
\vspace{-.1in}\eeq
where $\sum \pi_h = 1$.
The normal mixture model in (\ref{eq:DPM_model}) is assumed to be a Dirichlet process \citep{Ferguson73,Ishwaran01,LidHjort10}.  That is, the mixture probabilities follow a stick-breaking distribution \citep{Sethuraman94}, $\bpi=[\pi_{1},\pi_{2},\dots]' \sim \mbox{SB}(\varpi)$, or
\vspace{-.1in}\beq
\pi_{h} = v_h \prod_{g=1}^{h-1} (1-v_g)
\label{eq:v_m}
\vspace{-.15in}\eeq
where $v_h \stackrel{iid}{\sim} \mbox{Beta}(1,\varpi)$, $m=1,2,\dots$ A further hyper-prior is assumed on $\varpi$, i.e.,
\vspace{-.15in}\beq
\varpi  \sim \mbox{Gamma}(A_\varpi, B_\varpi).
\label{eq:hyper_gamma}
\vspace{-.18in}\eeq

It would be natural to assume remaining parameters $\bmu_h$ and $\bQ_h$ in (\ref{eq:DPM_model}) follow {\em iid} from the prior distribution defined in (\ref{eq:Sigma_prior}) and (\ref{eq:mu_prior}).  However, due to the large number of parameters, it is well known that mixture distributions do not perform well in moderate to high dimensions \citep{Raftery2006,Storlie17b}.  Thus, a sparse (s)DPM is constructed here as was done for the purpose of clustering with variable selection in \cite{Storlie17b}.

The description of the sDPM is facilitated by including a further latent variable $\phi$ to denote which mixture component produced $\bx^*$, i.e., $\Pr(\phi=h)=\pi_h$, and
\vspace{-.16in}\bdm
[\bx^* \mid \phi=h] \sim \cN(\bmu_h, \bQ^{-1}_h), 
\vspace{-.16in}\edm
The sDPM assumes that there are a subset of the variables $\bx_2^*$ that are {\em non-informative} in the sense that they are independent of $\phi$, conditional on the values of the other variables.  Partition the vector $\bx^* = [{\bx_1^*}' , {\bx_2^*}']'$ so that $\bx_1^*$ contains the {\em informative} variables that are affected by $\phi$ and $\bx_2^*$ are {\em non-informative} variables

Extending the notation from (\ref{eq:Sigma_prior}), define $\tbmu_h$ and $\tbQ_h$ so that $\mu_h = \tbD_h^{-1} \tbmu_h$ and $\bQ_h=\tbD_h \tbQ_h \tbD_h$.
Using the canonical parameterization of the Gaussian, $(\tbb_h, \tbQ_h)$, where $\tbb_h = \tbQ_h \tbmu_h$, the component parameters $\tbb_h$ and $\tbQ_h$ are partitioned accordingly, i.e.,
\vspace{-.08in}\bdm
\tbb_h= \left(
\begin{array}{c}
  \tbb_{h1}\\[-.05in]
  \tbb_{2}
\end{array}
\right), \;\;
\tbQ_h = \left(
\begin{array}{cc}
  \tbQ_{h11} & \tbQ_{12}\\[-.05in]
  \tbQ_{21} & \tbQ_{22}
\end{array}
\right),
\vspace{-.08in}\edm
where $(\tbb_{2}, \tbQ_{21}, \tbQ_{22})$ are constant across all components.  Under this parameterization, the non-informative variables $\tbx^*_2$ are independent of $\phi$, conditional on the values of the other variables \citep{Storlie17b}.  Inference into which variables are informative for determining component (cluster) membership may be of interest in its own right.  However, for the missing data regression problem the advantage gained is primarily improved density estimation due to the dimension reduction.

Let the informative variables for the sDPM be represented by the {\em model} $\bgamma$, a vector of binary values such that $\{ \tbx^*_j : \gamma_j=1\}$ is the set of informative variables and $\gamma_j=0$ for the non-informative variables.  A priori it is assumed that $\Pr(\gamma_j=1)=\rho_j$.  The same conjugate prior as that used in \cite{Storlie17b} is assumed on $(\tbb_h, \tbQ_h)$.
Let $\bPsi$ be a $p \times p$ positive definite matrix, partitioned just as $\tbQ_h$, and for a given $\bgamma$, the following prior distribution is assumed for $(\tbb_{2}, \tbQ_{21}, \tbQ_{22})$,
\vspace{-.15in}\beq
\begin{array}{rcl}
  \tbQ_{22} & \sim & \cW\left(\bPsi_{22 \mid 1}^{-1}, \eta\right), \\
  \tbQ_{21} \mid \tbQ_{22} & \sim & \cM\cN\left(-\tbQ_{22}\bPsi_{21}\bPsi_{11}^{-1} \; , \; \tbQ_{22} \; , \; \bPsi_{11}^{-1} \right), \\
  \tbb_2 \mid \tbQ_{22}   & \sim & \cN\left(\bzero, \frac{1}{\varphi}\bQ_{22}\right),
\end{array}
\label{eq:comp_prior_1}
\vspace{-.15in}\eeq
where $\cM\cN$ denotes the matrix normal distribution.
Further, assume the following distribution imposed on $(\tbb_{h1}, \tbQ_{h11})$, conditional on $(\tbb_{2}, \tbQ_{21}, \tbQ_{22})$,
\vspace{-.15in}\beq
\begin{array}{rcl}
  \tbSigma_{h11} & \stackrel{iid}{\sim} & \cI\cW \left(\bPsi_{11}, \eta - p_2^*\right), \\
  \tbmu_{h1} \mid \tbSigma_{h11}  & \stackrel{ind}{\sim} & \cN\left(\bzero, \frac{1}{\varphi}\tbSigma_{h11}\right),
\end{array}
\label{eq:comp_prior_2}
\vspace{-.1in}\eeq
via the relations $\tbb_h = \tbQ_h \tbmu_h$ and $\tbQ_h=\tbSigma_h^{-1}$, where $p_2^*=\sum I_{\{\gamma_j=0\}}$, $\cI\cW$ denotes the inverse-Wishart distribution, and $(\tbmu_{h1}, \tbSigma_{h11})$ are independent of $(\tbb_{2}, \tbQ_{21}, \tbQ_{22})$.

In mixture models it can be useful to impose a further hyper-priors on $\varphi$, $\eta$, and $\bPsi$ in (\ref{eq:comp_prior_1}) and (\ref{eq:comp_prior_2}) to provide less variable estimation of $\bmu_h$ and $\bSigma_h$ for classes $h$ that do not have many observations.  Therefore it is assumed that
\vspace{-.18in}\beq
\begin{array}{rcl}
  \varphi & \sim & \mbox{Gamma}(A_\varphi, B_\varphi), \\[-.06in]
  \eta & \sim & (p + 1) +\mbox{Gamma}(A_\eta, B_\eta), \\[-.06in]
  \bPsi & = & \psi \bI \\[-.06in]
  \psi & \sim & \mbox{Gamma}(A_\psi,B_\psi).
\end{array}
\label{eq:hyper_priors}
\vspace{-.12in}\eeq
For simplicity a mean $\bzero$ normal distribution is assumed for the prior of $\bmu_h$, which is reasonable if the $x_j$ are centered prior to analysis.  In addition, there are known issues with the use of a Wishart prior on variables of differing scale.  To alleviate this issue, we recommend first standardizing the columns of the data to have empirical mean zero and variance one prior to analysis, then setting $A_\varphi , B_\varphi , A_\eta , B_\eta , A_\psi$, and $B_\psi$ all equal 1.

\begin{figure}[t!]
\vspace{-.09in}\centering
\caption{Comparison of sDPM and MVN models on the bivariate marginal distribution of the variables {\em spo2} and log({\em troponin}+1). (a) Sample of 2000 observations from BPR data; dashed lines indicate hard boundaries, i.e., {\em spo2} cannot be above 100 percent, and {\em troponin} cannot be negative.  (b) Realization of 2000 observations randomly generated from the MVN model with no censoring  (c) Realization of 2000 observations from the MVN model with censoring. (d) Realization of 2000 observations from the sDPM model with censoring; parameters for plots (b)-(d) obtained from the last of 10,000 posterior samples fit to the full BPR data set.}
\vspace{-.07in}
 \begin{subfigure}[b]{.425\textwidth}
      \centering
      \caption{Observed Data}
      \vspace{-.13in}
      \includegraphics[width=.93\textwidth, height=.25\textheight]{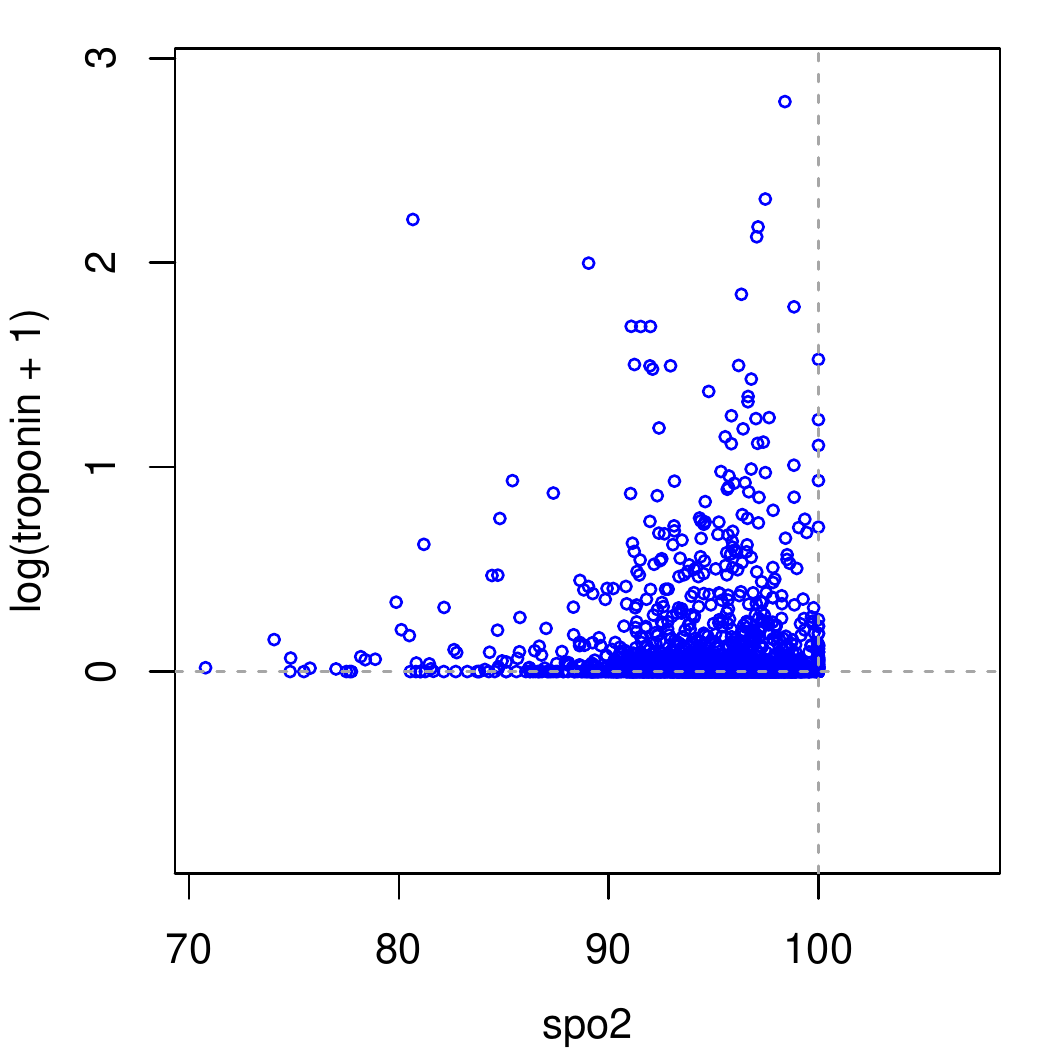}
    \end{subfigure}
    \begin{subfigure}[b]{.425\textwidth}
      \centering
      \caption{MVN model}
      \vspace{-.13in}
      \includegraphics[width=.93\textwidth, height=.25\textheight]{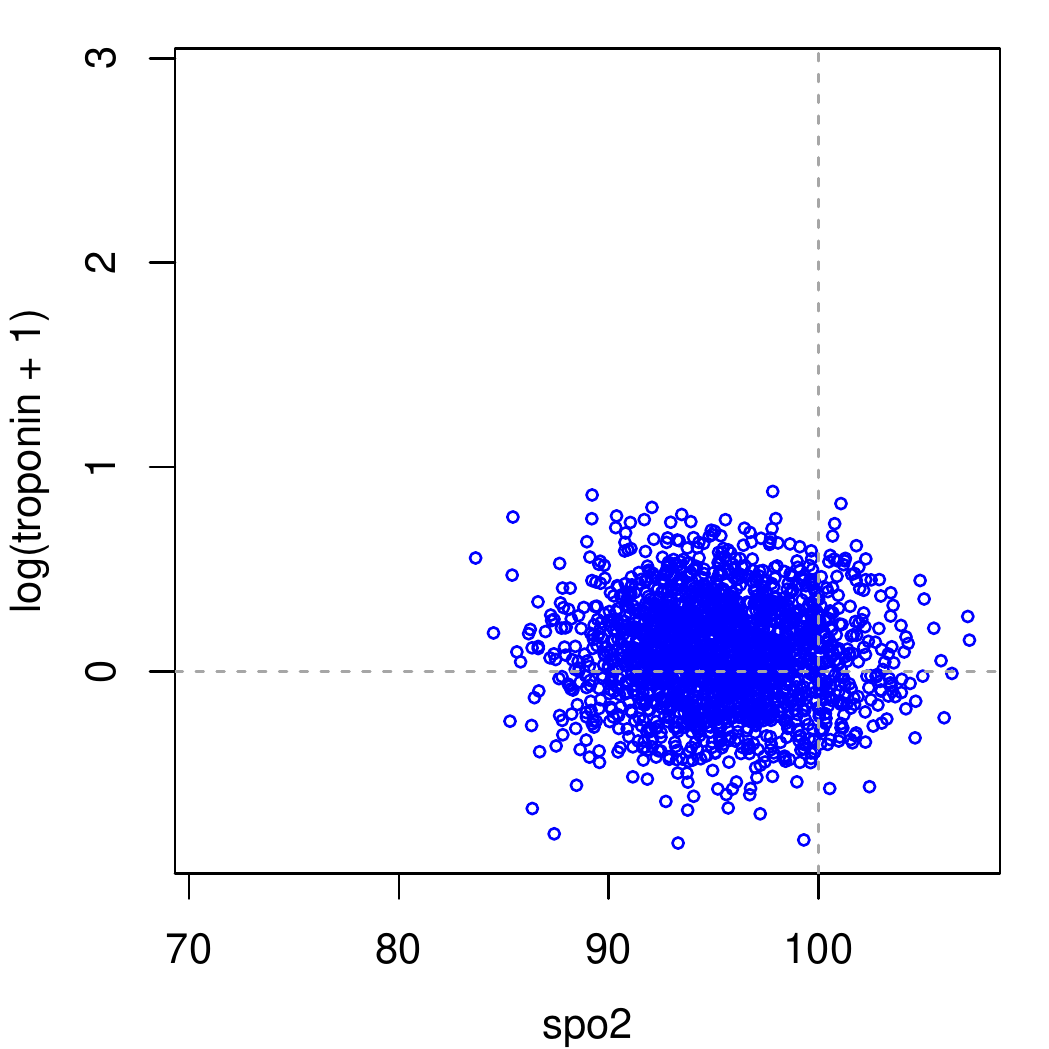}
    \end{subfigure}
    \begin{subfigure}[b]{.425\textwidth}
      \centering
      \vspace{.1in}
      \caption{MVN model with censoring}
      \vspace{-.13in}
      \includegraphics[width=.93\textwidth, height=.25\textheight]{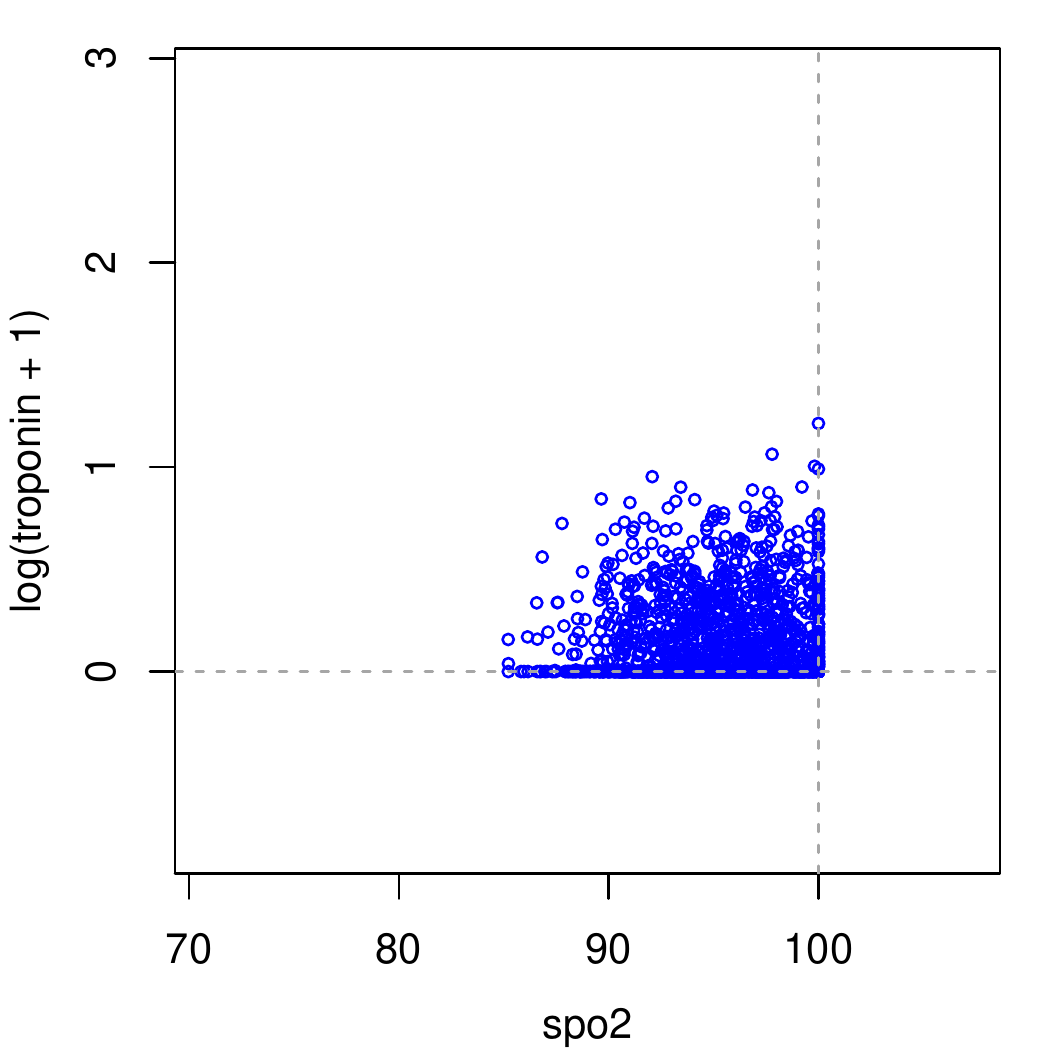}
    \end{subfigure}
    \begin{subfigure}[b]{.425\textwidth}
      \centering
      \vspace{.1in}
      \caption{sDPM model with censoring}
      \vspace{-.13in}
      \includegraphics[width=.93\textwidth, height=.25\textheight]{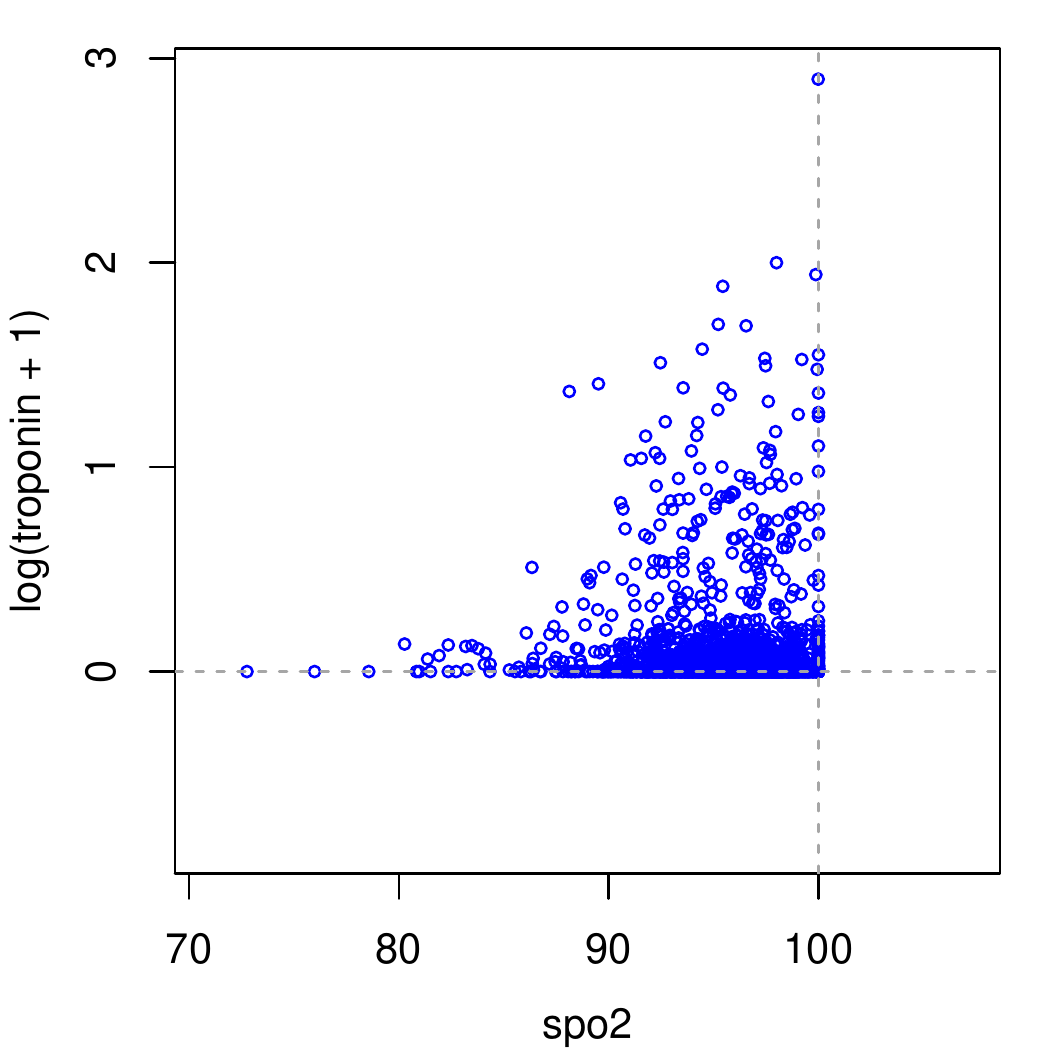}
    \end{subfigure}
\label{fig:DPM_MVN}
\vspace{-.08in}\end{figure}

Figure~\ref{fig:DPM_MVN} shows a comparison of the MVN without censoring, MVN with censoring, and sDPM for $\bx^*$ when fit to the BPR data using the MCMC procedure described in Section~\ref{sec:MCMC}.  Figure~\ref{fig:DPM_MVN}(a) is a bivariate scatter plot of the variables {\em spo2} and log({\em lab.value.troponin}+1) from the BPR data (a sample of size 2,000 from the 34,923 observations was used for better display).  The plot in Figure~\ref{fig:DPM_MVN}(b) is the result of fitting the model assuming a MVN distribution for $\bx^*$.  The plot provides a random generation of 2,000 points from the respective bivariate marginal distribution resulting from the fitted normal distribution.  The realization uses a draw from the posterior distribution for $\bmu$ and $\bQ$ to define this bivariate distribution, but the fitted model included all 171 predictors.  There is clearly a failure to capture the distributional structure in any meaningful way by the MVN.  Many values ($\sim$72\%) are missing for {\em lab.value.troponin}, and using a MVN would produce very misleading imputations during the MCMC.  Figure~\ref{fig:DPM_MVN}(c) provides a random realization from the MVN approach described in Section~\ref{sec:model_descr} that accounts for censoring/limits for the variables.  This is a much better approximation to the underlying distribution of the data in Figure~\ref{fig:DPM_MVN}(a), but the distribution of the actual data has much heavier tails.  Finally, Figure~\ref{fig:DPM_MVN}(d) shows a realization from the fitted sDPM model, which produces a much better approximation to the underlying distribution in this case.

\vspace{-.15in}
\subsection{BPR Regression Model}
\vspace{-.07in}
\label{sec:reg_model}

Any regression model can be used in place of (\ref{eq:weibull_reg}) and the missing data framework for $\bx$ remains unchanged.
However, for the BPR problem the time to event data $y$ are assumed to follow a Weibull model,
\vspace{-.15in}\begin{eqnarray}
y_i & \!\!\stackrel{iid}{\sim} \!\!& \mbox{Weibull}(\lambda(\bx_i), \kappa), \label{eq:weibull_reg} \\[-.0in]
\log \lambda(\bx_i) & \!\!=\!\! & \beta_{0,1} + \sum_{j=1}^r \sum_{k=1}^{L_j-1} \beta_{j,k} I_{\{x_{i,j}=k\}} + \!\sum_{j=r+1}^p \beta_{j,1} x_{i,j} + \!\sum_{j=p+1}^{p+K}\!\! \beta_{j,1} \:g_{j-p}(\bx_i), \label{eq:weibull_reg2} \\[-.4in]
\nonumber
\end{eqnarray}
where the first $r$ of the $x_j$ are categorical with $L_j-1$ levels, and $g_k$, $k=1,\dots,K$ are optional functions of the $x_{i,j}$ to allow for a set of possible interactions, for example.  A complete list of such interactions used in the BPR problem is provided in the Supplementary Material.  As in \cite{Storlie12a} a point mass prior is imposed on $\bbeta = [\beta_{0,1},\beta_{1,1}, \dots, \beta_{1,L_1-1},\dots,\beta_{p+K,1}]'$ to facilitate (grouped) variable selection,
\vspace{-.15in}\begin{eqnarray}
\beta_{j,k} & = & \delta_j \xi_{j,k}, \label{eq:beta_prior} \\[-.05in]
\delta_j & \stackrel{ind}{\sim} & \mbox{Bernoulli}(\rho_j), \nonumber \\[-.05in]
\xi_{j,k} & \stackrel{iid}{\sim} & \mbox{N}(0, \tau^2), \nonumber \\[-.05in]
\tau^{-2} & \sim & \mbox{Gamma}(A_\tau,B_\tau). \nonumber \\[-.5in]
\nonumber
\end{eqnarray}
The $\delta_j$ are indicator variables that permit variable selection for the $j\th$ variable (i.e., if $\delta_l=0$ then all $\beta_{j,k}=0$ for the $j\th$ variable).  Conditional on $\delta_j=1$ the $\beta_{j,k}$ for the $j\th$ variable are independent and normally distributed.  It is conventional to assume $\rho_0=1$ so that $\delta_0=1$ and the intercept is always in the model.  Finally, it is assumed 
\vspace{-.15in}\begin{eqnarray}
\kappa & \sim & \mbox{Gamma}(A_\kappa,B_\kappa), \nonumber \\[-.05in]
\bx^*_i & \stackrel{iid}{\sim} & \mbox{DPM} \mbox{, as in (\ref{eq:DPM_model})}. \nonumber \\[-.5in]
\nonumber
\end{eqnarray}

\vspace{-.25in}
\subsection{MCMC Algorithm}
\vspace{-.1in}
\label{sec:MCMC}

Complete MCMC details are provided in the Supplementary Material.  However, an overview is provided here to illustrate the main idea.  The MCMC routine is a typical hybrid Gibbs, Metropolis Hastings (MH) sampling scheme (e.g., see \cite{Kruschke2014}).
The complete list of parameters in the model described in (\ref{eq:reg_model}) and (\ref{eq:DPM_model}) are
\vspace{-.20in}\beq
\Theta = \left\{
\btheta,
\left\{\pi_{h}, \bmu_{h}, \bQ_{h}\right\}_{h=1}^H,
\varphi, \eta, \psi, 
\bX^*,
\bphi,
\bgamma
\right\},
\label{eq:param_list}
\vspace{-.21in}\eeq
where $\btheta$ are the regression model parameters ($\btheta=\{\bbeta,\tau^2,\kappa\}$ for this application) and $\bX^* = \{\bx_1^*,\dots,\bx_n^* \}'$ where $\bx_i^*$ are the values of latent vector $\bx^*$ for the $i\th$ observation.
The number of components in the stick-breaking model for $\pi_{h}$ was truncated at a finite value $H$, i.e., $h=1,\dots,H$, as in \cite{Ishwaran01}.  The value of $H$ was set such that roughly half of the $\pi_{h}$ values were negligible ($H = 40$ for the BPR problem) to ensure a good approximation to the fully nonparametric DPM.

With the inclusion of the $\bphi$, conditional on the {\em rest}, $\pi_h$ has a conjugate update.  The $\bgamma$ must be sampled together with the $\tbmu_h$ and $\tbQ_h$ because their dimensionality changes with the value of $\bgamma$, which requires a reversible jump MCMC update \citep{Green1995,Green2009}.  This is streamlined by the fact that there is a conjugate update for the $\tbmu_h$ and $\tbQ_h$ conditional on $\bgamma$, thus providing a convenient means of producing reasonable proposals for the the $\tbmu_h$ and $\tbQ_h$, independent of the current state; this makes the Jacobian for the dimension matching transformation trivially equal to one.  As stated previously, the likelihood identified parameters $\bmu_h$ and $\bQ_h$ are simply computed via deterministic relations after the update of $\tbmu_h$ and $\tbQ_h$.  The latent indicator $\bphi$ can be sampled directly from its full conditional distribution.  The scalar parameters $\varphi$, $\eta$, and $\psi$ do not have convenient full conditional forms and are thus updated with a typical random walk MH on log scale.

The rows of $\bX^*$, $\bx^*_i$, $i=1,\dots,n$ can be updated independently and in parallel.  If $x_{i,j}$ is {\bf not} missing, then the corresponding element(s) of $\bx^*_i$ has a simple truncated normal update for $j \leq r$, and if $j>r$ then it is fixed and equal to $x_{i,j}$.  However, when $x_{i,j}$ is missing, there is no convenient distributional form for updating the the corresponding elements of $\bx^*_i$ due to the dependence on the response $y_i$.  If, on the other hand, there were no $y_i$ included in the {\em rest}, then they would have normal updates, conditional on the remaining elements of $\bx_i^*$.  Thus, an effective strategy is to draw a proposal from the closed form update as if there were no $y_i$ and use this draw as a proposal in a MH step.  This approach requires no tuning and resulted in high acceptance rates for all $x^*_{i,j}$ (mean acceptance was 76\%, lowest was 46\%) in the BPR analysis and similarly high acceptance rates in all of our simulation study cases.

Updates for $\btheta$ are regression model dependent.  However, whatever the typical update is for the given model in the complete data case would apply, since the current value for $X_{\miss}$ is treated as given.  For example, if the regression model is a normal error, linear model, then the $\btheta$ will have a usual conjugate normal (and inverse gamma) update.  In the case of the regression model in (\ref{eq:weibull_reg}) there is no such convenient update.  However, a suitable transformation of the response allows for a conjugate normal approximation to serve as a good proposal for $\bbeta_j$ to be accepted/rejected in an MH step under the Weibull likelihood. Again, this approach requires no tuning and resulted in high acceptance rates for all $\bbeta_j$ (lowest acceptance rate of 55\% in the BPR analysis with $\sim$200 elements in $\bbeta$).   The $\kappa$ parameter is updated with a typical random walk MH on log scale and thus $\kappa$, $\varphi$, $\eta$, and $\psi$ are the only updates for the BPR application that require specification of a tuning parameter.  However, they are a scalars so this is fairly straight-forward.

\vspace{-.2in}
\section{Simulation Results}
\vspace{-.09in}
\label{sec:sims}

In this section the performance of the proposed approach for BPR is compared to several other methods on three simulation cases, similar in nature to the BPR problem.   Each case has 50 predictors, of which the first 25 are discrete binary predictors, the last 25 are continuous, and there are no additional functions $g_k$, included, i.e., $K=0$.  The simulated training data has 2,500 observations, generated from either a MVN or DPM model for $\bx^*$, as described below, and $y_i \stackrel{ind}{\sim}\mbox{Weibull}(\lambda(\bx_i), 2)$, with
\vspace{-.1in}\beq
\bbeta = [\underbrace{0}_\text{intercept},\underbrace{\:1.0\:,\:0.5\:,\:-0.5\:,\:0.2\:,\:0\:,\:\:\dots\:,\:0}_\text{coefficients on discrete $x_1,\dots,x_{25}$}\:,\:\underbrace{1.5\:,\:0.5\:,\:-0.5\:,\:0.2\:,\:0\:,\:\dots\:,\:0}_\text{coefficients on continuous $x_{26},\dots,x_{50}$} \;\;]'.
\label{eq:sim_beta}
\vspace{-.1in}\eeq
The $y_i$ were then right censored at 0.20, resulting in $\sim$80\% of the $y_i$ being censored in all cases.  The three simulation cases are described below:
\begin{description}
\small  \singlespacing
\item[Case 1:]{$\bx^* \sim \mbox{N}(\bzero, \bQ^{-1})$, where $\bQ$ was randomly generated prior to each data realization according to  $\tbQ \sim \mbox{Wishart}(55, 0.25\bI)$, and $\bQ=\tbD \tbQ \tbD$ as in (\ref{eq:Sigma_prior}).  All continuous variables were then standardized to mean 0, standard deviation 0.5 (to make coefficients comparable between binary and continuous variables) prior to generating the $y_i$ from the relation in (\ref{eq:sim_beta}).  All predictors are missing completely at random with a missing rate of 50\% for $x_1$ and $x_{26}$, and a missing rate of 25\% for all other predictors.}\\[-.25in]
\item[Case 2:]{Same as Case 1 except that the predictors $x_1, x_2, x_3, x_{26}, x_{27}, x_{28},$ are missing not at random with probability determined via a probit relationship to their underlying values to provide an overall missing rate of 50\% for $x_1,x_{26}$ and 25\% for  $x_2, x_3, x_{27}, x_{28}$.  All other predictors are missing completely at random with a missing rate of 25\%.}\\[-.25in]
\item[Case 3:]{$\bx^* \sim \mbox{sDPM}$, where $\bpi \sim \mbox{SB}(0.5)$, to provide an average of about six non-negligible (i.e., those with $\pi_h>0.01$) components.  The $\tbmu_h$, $\tbQ_h$ were generated as in (\ref{eq:comp_prior_1})~and~(\ref{eq:comp_prior_2}) with $\bPsi=0.25\bI$, $\eta=55$, and $\varphi=1$.  As in Case 1, all predictors are missing completely at random with a missing rate of 50\% for $x_1$ and $x_{26}$, and a missing rate of 25\% for all other predictors.}\\[-.25in]
\item[Case 4:]{Same as Case 3 except that the predictors $x_1, x_2, x_3, x_{26}, x_{27}, x_{28}$ are missing not at random in the same manner as that in Case 2.}\\[-.25in]
\item[Case 5:]{$\bx$ is constructed by randomly selecting $p=50$ of the 171 predictors in the BPR dataset.  The $n=2,500$ observations of these 50 predictors are generated by sampling from the empirical distribution of the BPR data.  However, the empirical distribution was necessarily conditioned/restricted to those observations where $x_j$ was observed (i.e., not missing) for the informative predictors (i.e., $j=1,2,3,26,27,28$) so that the regression function could be computed to obtain $\lambda(\bx_i)$ for each observation.  Once used to calculate $\lambda(\bx_i)$ the informative $x_j$ were made missing completely at random with a missing rate of 50\% for $x_1$ and $x_{26}$, and a missing rate of 25\% for $x_2,x_3,x_{27},x_{28}$.
  This data generating process resulted in an average of about $\sim \!22$ discrete and $\sim \!28$ continuous variables.} \\[-.25in]
\item[Case 6:]{Same as Case 5 except that the predictors $x_1, x_2, x_3, x_{26}, x_{27}, x_{28}$ are missing not at random in the same manner as that in Case 2.}\\[-.25in]
\item[Case 7:]{Same as Case 5 except that the predictors $x_2$ and $x_{27}$ are completely removed from the training data and are thus not available for  model fitting to mimic the reality that not {\em all} relevant predictors will typically be available.}\\[-.25in]
\item[Case 8:]{Same as Case 7 except that the predictors $x_1, x_2, x_3, x_{26}, x_{27}, x_{28}$ are missing not at random in the same manner as that in Case 2.}\\[-.25in]
\end{description}

\noindent The following methods were used to fit a model to the simulated data:
\begin{description}
\small  \singlespacing
\item[MVN-MAR:] $\bx^* \sim \cN(\bmu, \bSigma)$; {\em no} indicators of missingness included in the $\bx$ model.\\[-.25in]
\item[MVN-MNAR:] $\bx^* \sim \cN(\bmu, \bSigma)$; latent variable for binary missingness indicator for each predictor included in the model for $\bx^*$.\\[-.25in]
\item[sDPM-MAR:] $\bx^* \sim sDPM$; {\em no} indicators of missingness included in the $\bx$ model.\\[-.25in]
\item[sDPM-MNAR:] $\bx^* \sim sDPM$; latent variable for binary missingness indicator included for each predictor included in the model for $\bx^*$.\\[-.25in]
\item[RFEN-MAR:] Random Forest Imputation to fill in missing via the \verb1missForest1 package in R, then Elastic Net (EN) Cox Regression via the \verb1glmnet1 package in R.\\[-.25in]
\item[GBM:] Gradient Boosting Machine Cox regression via the R package \verb1gbm1, with depth, shrinkage and number of trees chosen via 10-fold cross validation.\\[-.25in]
\item[CPH-SW:] Include a ``missing'' level for discrete predictors, and include a missing indicator for continuous predictors; Stepwise regression via the \verb1step1 function of a Cox proportional hazards model via the \verb1survival1 package in R.\\[-.25in]
\end{description}

A total of 100 data sets were generated from each simulation case and each of the methods above was fit to each data set.  An independent test set of 1,000 observations was then generated for each data set and each method was used to provide predictions $\hlambda(\bx_i)$ of the true relative risk $\lambda(\bx_i)$ for each observation in the test set.  The predictive performance is summarized in terms of (i) the concordance (or C-statistic) of the predicted relative risk to the test set event times, and (ii) the $R^2$ of the predicted log-risk to the {\em true} relative log-risk (i.e., the true risk is computed using the relation in (\ref{eq:sim_beta}) with all covariate values and coeeficients exactly known).  These measures are summarized by the mean of these quantities over the 100 data sets in the tables below.  The number of variables selected  ({\em Model Size}) and the proportion of variables correct ({\em PVC}$\:$), i.e., correctly included/excluded in the model, are summarized by their respective means as well.  The summary score for the {\em best} method for each summary is bolded along with that for any other method that was not statistically different (at a 0.05, unadjusted level of significance) from the {\em best} method on the basis of the 100 trials.  The estimated risk $\hlambda(\bx_i)$ for the proposed Bayesian approaches are taken to be the mean of 5,000 posterior sample predictions.
The {\em True} method is not a method at all, rather it is just a reference to display the best possible obtainable values for each summary if the missing predictors were imputed perfectly and the regression relation was known.

\renewcommand{\arraystretch}{.99} 
\begin{table}[t!]
  \vspace{-.08in}
\setlength\tabcolsep{.075in}
      \begin{widepage}
  \begin{center}
    \caption{Simulation Results.}
    \vspace{-.075in}
\label{tab:sim}
\footnotesize
  \label{tab:case1}
  \begin{tabular}{lcccccccccc}
    \hline \\[-.12in]
   \multirow{2}{*}{\bf Method} & &  \bf \!\!\!Concord & \bf Risk $R^2\!\!$  & $\!\!\!$\bf Size$\!\!\!$& \bf PVC & & \bf Concord & \bf Risk $R^2\!\!$  & $\!\!\!$\bf Size$\!\!\!$& \bf PVC\\[.05in]
\cline{3-6}
\cline{8-11} \\[-.07in]
& & \multicolumn{4}{c}{Case 1: $\bz\sim$MVN and MAR} &&  \multicolumn{4}{c}{Case 2: $\bz\sim$MVN and MNAR} \\[.03in]
\cline{1-1}\cline{3-6}\cline{8-11}
\!\! MVN-MAR \!\!\!& & {\bf 0.766} & {\bf 0.797} & {\bf  6.49} & {\bf 0.976} & & 0.743 & 0.704 &  6.78 & 0.956 \\
\!\! MVN-MNAR \!\!\!& & 0.765 & 0.792 & {\bf  6.50} & {\bf 0.974} & & {\bf 0.762} & {\bf 0.789} & {\bf  6.33} & {\bf 0.983} \\
\!\! sDPM-MAR \!\!\!& & {\bf 0.766} & {\bf 0.797} & {\bf  6.32} & {\bf 0.976} & & 0.743 & 0.705 &  6.84 & 0.957 \\
\!\! sDPM-MNAR \!\!\!& & 0.765 & 0.792 & {\bf  6.31} & {\bf 0.973} & & {\bf 0.762} & {\bf 0.790} & {\bf  6.31} & {\bf 0.983} \\
\!\! GBM \!\!\!& & 0.715 & 0.548 & 22.30 & 0.448 & & 0.736 & 0.637 & 20.90 & 0.495 \\
\!\! RFEN-MAR \!\!\!& & 0.744 & 0.690 & 12.40 & 0.744 & & 0.721 & 0.602 & 14.40 & 0.677 \\
\!\! CPH-SW \!\!\!& & 0.727 & 0.616 & 16.70 & 0.609 & & 0.748 & 0.726 & 15.40 & 0.665 \\
\cline{1-1}\cline{3-6}\cline{8-11}
\!\! TRUE \!\!\!& & 0.812 & 1.000 &  6.00 & 1.000 & & 0.810 & 1.000 &  6.00 & 1.000 \\
\cline{1-1}\cline{3-6}\cline{8-11}\\[-.07in]
& & \multicolumn{4}{c}{Case 3: $\bz\sim$sDPM and MAR} &&  \multicolumn{4}{c}{Case 4: $\bz\sim$sDPM and MNAR} \\[.03in]
\cline{1-1}\cline{3-6}\cline{8-11}
\!\! MVN-MAR \!\!\!& & 0.747 & 0.706 & {\bf  6.60} & {\bf 0.975} & & 0.702 & 0.529 &  9.35 & 0.849 \\
\!\! MVN-MNAR \!\!\!& & 0.745 & 0.697 & {\bf  6.68} & {\bf 0.971} & & 0.736 & 0.663 &  8.94 & 0.878 \\
\!\! sDPM-MAR \!\!\!& & {\bf 0.779} & {\bf 0.825} & {\bf  6.51} & {\bf 0.982} & & 0.770 & 0.803 & {\bf  7.10} & {\bf 0.953} \\
\!\! sDPM-MNAR \!\!\!& & 0.777 & 0.814 & {\bf  6.60} & {\bf 0.978} & & {\bf 0.795} & {\bf 0.903} & {\bf  6.66} & {\bf 0.968} \\
\!\! GBM \!\!\!& & 0.719 & 0.578 & 21.10 & 0.493 & & 0.769 & 0.752 & 18.60 & 0.577 \\
\!\! RFEN-MAR \!\!\!& & 0.755 & 0.733 & 11.80 & 0.799 & & 0.735 & 0.661 & 14.00 & 0.713 \\
\!\! CPH-SW \!\!\!& & 0.703 & 0.528 & 15.80 & 0.650 & & 0.772 & 0.813 & 12.90 & 0.767 \\
\cline{1-1}\cline{3-6}\cline{8-11}
\!\! TRUE \!\!\!& & 0.822 & 1.000 &  6.00 & 1.000 & & 0.822 & 1.000 &  6.00 & 1.000 \\
\cline{1-1}\cline{3-6}\cline{8-11}\\[-.07in]
& & \multicolumn{4}{c}{Case 5: $\bz\sim$BPR and MAR} &&  \multicolumn{4}{c}{Case 6: $\bz\sim$BPR and MNAR} \\[.03in]
\cline{1-1}\cline{3-6}\cline{8-11}
\!\! MVN-MAR \!\!\!& & 0.761 & 0.728 & {\bf  6.71} & {\bf 0.962} & & 0.760 & 0.714 & {\bf  7.01} & {\bf 0.952} \\
\!\! MVN-MNAR \!\!\!& & 0.760 & 0.734 & {\bf  6.67} & {\bf 0.962} & & 0.760 & 0.733 & {\bf  6.86} & {\bf 0.954} \\
\!\! sDPM-MAR \!\!\!& & {\bf 0.764} & {\bf 0.753} & {\bf  6.53} & {\bf 0.971} & & 0.759 & 0.721 & {\bf  6.89} & {\bf 0.953} \\
\!\! sDPM-MNAR \!\!\!& & 0.761 & 0.718 & {\bf  6.73} & {\bf 0.962} & & {\bf 0.762} & {\bf 0.745} & {\bf  6.67} & {\bf 0.962} \\
\!\! GBM \!\!\!& & 0.714 & 0.524 & 23.70 & 0.396 & & 0.710 & 0.508 & 23.40 & 0.406 \\
\!\! RFEN-MAR \!\!\!& & 0.761 & 0.737 & 11.00 & 0.816 & & 0.757 & 0.724 & 10.70 & 0.825 \\
\!\! CPH-SW \!\!\!& & 0.735 & 0.627 & 13.60 & 0.719 & & 0.735 & 0.626 & 13.90 & 0.712 \\
\cline{1-1}\cline{3-6}\cline{8-11}
\!\! TRUE \!\!\!& & 0.823 & 1.000 &  6.00 & 1.000 & & 0.823 & 1.000 &  6.00 & 1.000 \\
\cline{1-1}\cline{3-6}\cline{8-11}\\[-.07in]
& & \multicolumn{4}{c}{$\!\!\!$Case 7: $\!\bz\sim$BPR, MAR, $x_2,x_{27}$ removed$\!\!\!$} &&  \multicolumn{4}{c}{$\!\!\!$Case 8: $\!\bz\sim$BPR, MNAR, $x_2,x_{27}$ removed$\!\!\!$} \\[.03in]
\cline{1-1}\cline{3-6}\cline{8-11}
\!\! MVN-MAR \!\!\!& & 0.735 & 0.612 & {\bf  6.67} & {\bf 0.891} & & 0.737 & 0.639 & {\bf  7.02} & {\bf 0.873} \\
\!\! MVN-MNAR \!\!\!& & 0.742 & 0.654 & {\bf  6.40} & {\bf 0.900} & & 0.744 & 0.678 & {\bf  7.19} & {\bf 0.870} \\
\!\! sDPM-MAR \!\!\!& & {\bf 0.746} & {\bf 0.670} & {\bf  6.25} & {\bf 0.909} & & 0.740 & 0.650 & {\bf  6.96} & {\bf 0.879} \\
\!\! sDPM-MNAR \!\!\!& & 0.734 & 0.599 & {\bf  6.62} & {\bf 0.894} & & {\bf 0.747} & {\bf 0.693} & {\bf  6.72} & {\bf 0.881} \\
\!\! GBM \!\!\!& & 0.700 & 0.472 & 22.00 & 0.354 & & 0.700 & 0.474 & 22.50 & 0.329 \\
\!\! RFEN-MAR \!\!\!& & {\bf 0.743} & {\bf 0.664} &  9.73 & 0.780 & & 0.743 & 0.676 & 10.50 & 0.753 \\
\!\! CPH-SW \!\!\!& & 0.720 & 0.561 & 12.50 & 0.677 & & 0.722 & 0.578 & 12.50 & 0.676 \\
\cline{1-1}\cline{3-6}\cline{8-11}
\!\! TRUE \!\!\!& & 0.825 & 1.000 &  4.00 & 1.000 & & 0.819 & 1.000 &  4.00 & 1.000 \\
\cline{1-1}\cline{3-6}\cline{8-11}\\[-.07in]
  \end{tabular}
  \end{center}
      \end{widepage}
  \vspace{-.19in}
\end{table}

\begin{figure}[t!]
  \vspace{-.08in}
      \begin{widepage}
\centering
\caption{$R^2$ of predicted log-risk $\hat{\lambda}$ to the true value.}
    \begin{subfigure}[b]{0.286\textwidth}
      \centering
  \vspace{-.06in}
      \caption{Case 1$\!\!\!\!\!\!\!\!\!\!\!\!\!\!\!$}
  \vspace{-.1in}
      \includegraphics[width=.99\textwidth]{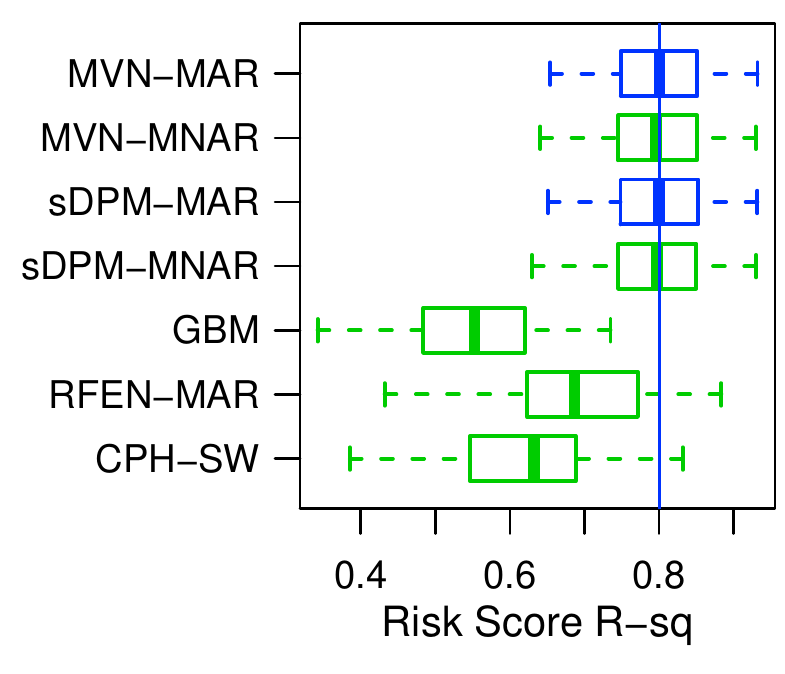}
    \end{subfigure}
    \begin{subfigure}[b]{0.200\textwidth}
      \centering
  \vspace{-.06in}
      \caption{Case 2}
  \vspace{-.1in}
      \includegraphics[width=.99\textwidth]{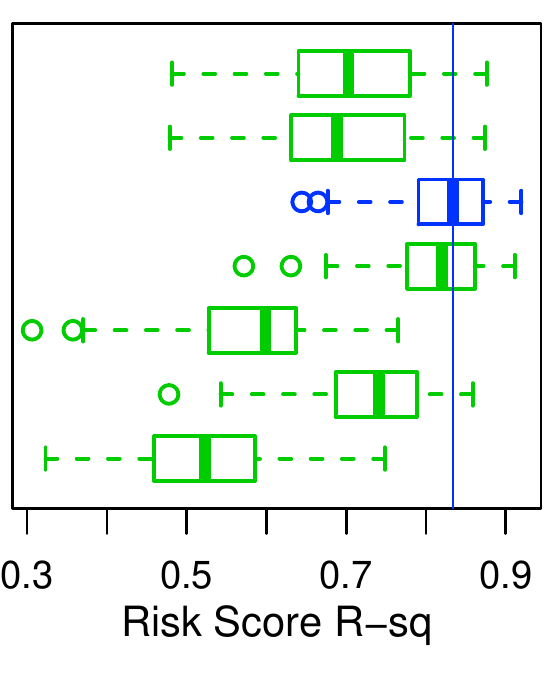}
    \end{subfigure}
    \begin{subfigure}[b]{0.200\textwidth}
      \centering
  \vspace{-.06in}
      \caption{Case 3}
  \vspace{-.1in}
      \includegraphics[width=.99\textwidth]{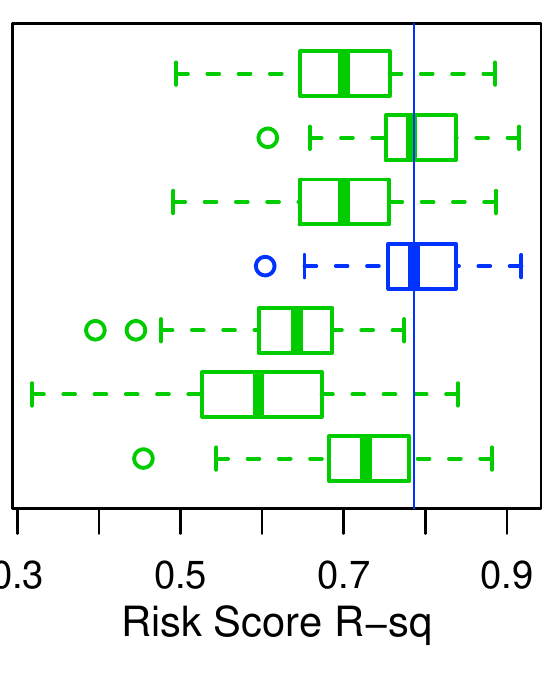}
    \end{subfigure}
    \begin{subfigure}[b]{0.286\textwidth}
      \centering
  \vspace{-.06in}
      \caption{Case 4$\;\;\;\;\;\;\;\;\;\;$}
  \vspace{-.1in}
      \includegraphics[width=0.99\textwidth]{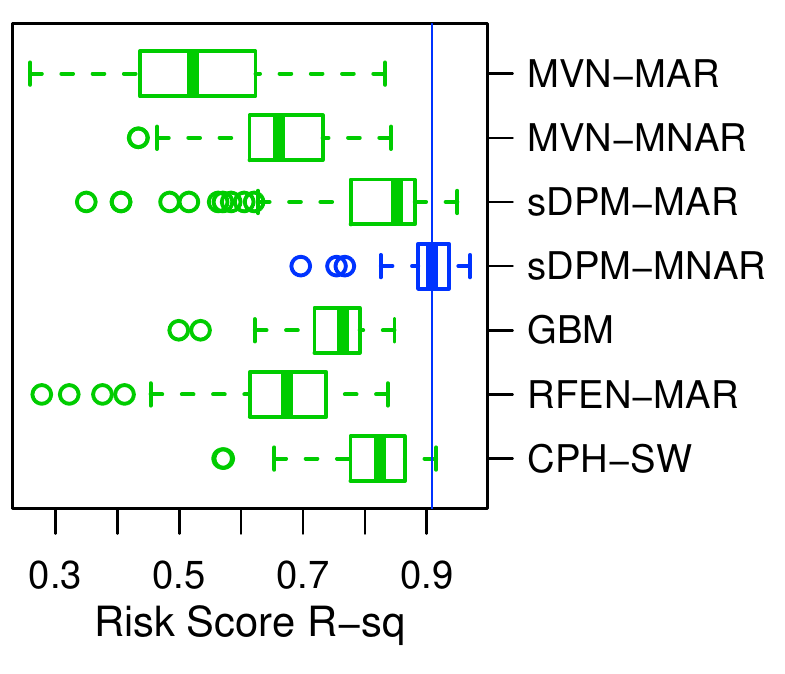}
    \end{subfigure}
    \begin{subfigure}[b]{0.286\textwidth}
      \centering
  \vspace{.05in}
      \caption{Case 5$\!\!\!\!\!\!\!\!\!\!\!\!\!\!\!$}
  \vspace{-.1in}
      \includegraphics[width=.99\textwidth]{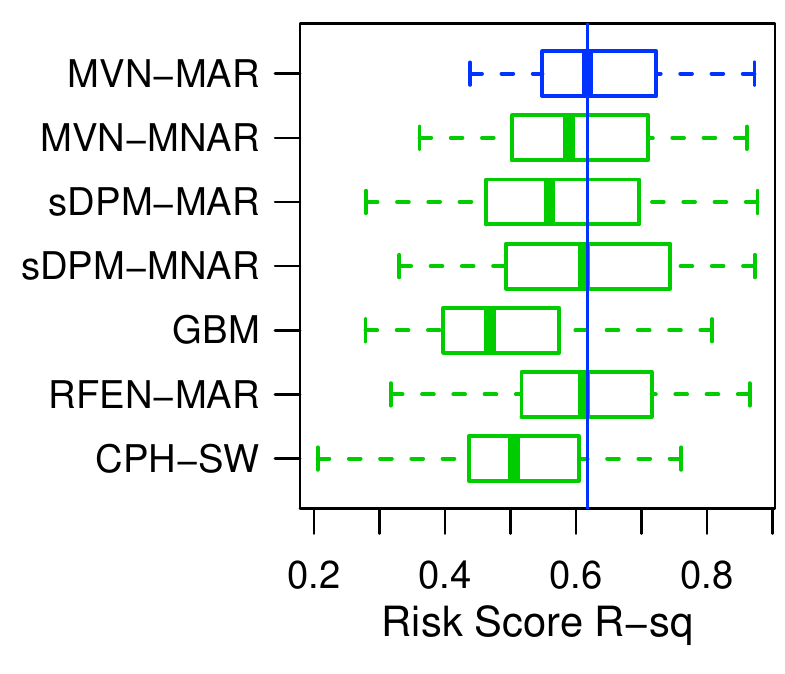}
    \end{subfigure}
    \begin{subfigure}[b]{0.200\textwidth}
      \centering
      \caption{Case 6}
  \vspace{-.1in}
      \includegraphics[width=.99\textwidth]{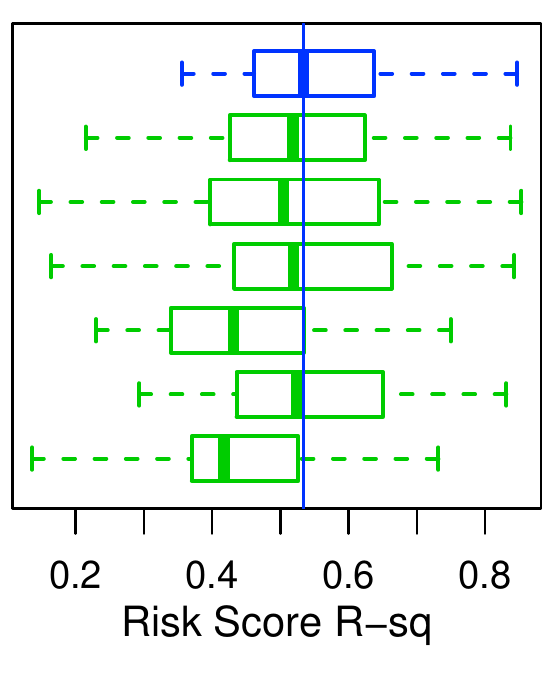}
    \end{subfigure}
    \begin{subfigure}[b]{0.200\textwidth}
      \centering
      \caption{Case 7}
  \vspace{-.1in}
      \includegraphics[width=.99\textwidth]{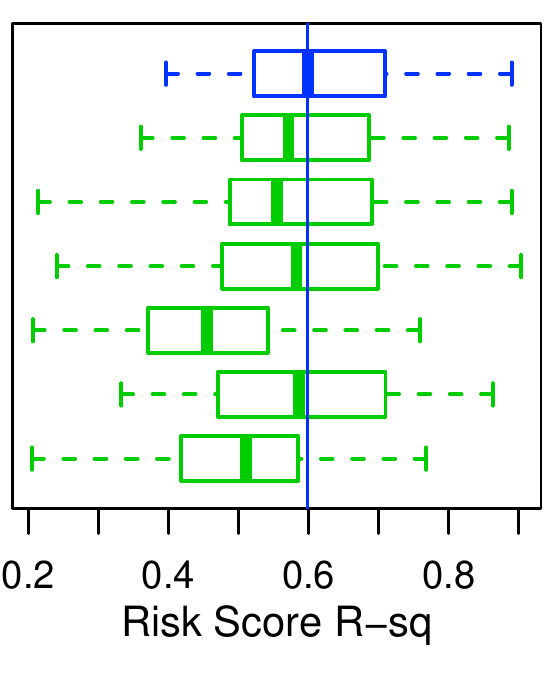}
    \end{subfigure}
    \begin{subfigure}[b]{0.286\textwidth}
      \centering
      \caption{Case 8$\;\;\;\;\;\;\;\;\;\;$}
  \vspace{-.1in}
      \includegraphics[width=0.99\textwidth]{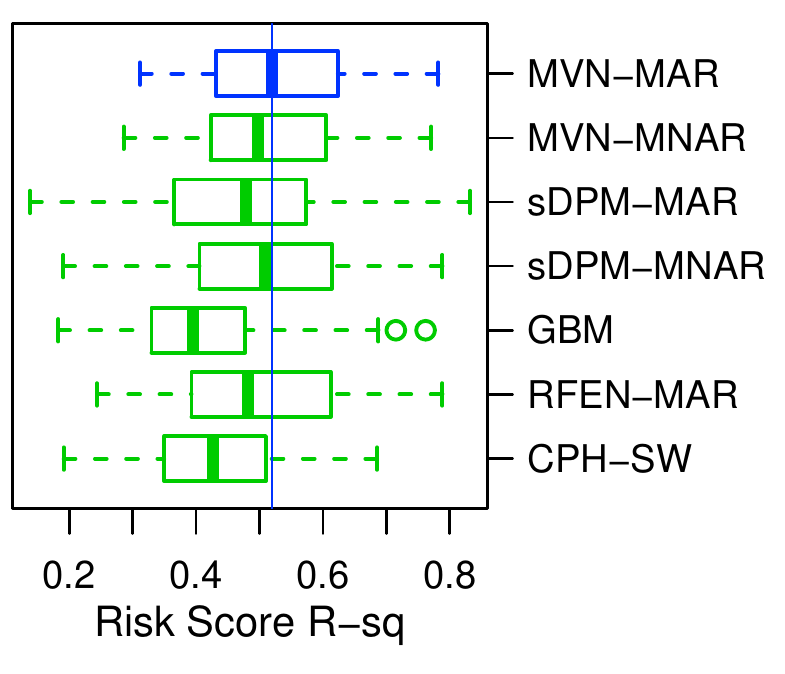}
    \end{subfigure}
\label{fig:risk_Rsq}
      \end{widepage}
  \vspace{-.25in}
\end{figure}

The simuation results from the eight cases are summarized in Table~\ref{tab:sim}.  Since these data were generated by the MVN-MAR model, the MVN-MAR approach has the best concordance and Risk $R^2$, and the largest proportion of correctly identified variables.  However, these results are not statistically different from those from sDPM-MAR, nor are they practically different from MVN-MNAR and sDPM-MNAR.  All four of these approaches are distinctly better than the others with respect to all measures.  Figure~\ref{fig:risk_Rsq} provides boxplots of the risk $R^2$ across the 100 simulatiuons for each of the methods and each simulation case.  Methods not significantly different from the {\em best} are in blue, others in green.
In Case 2 the model is generated by sDPM-MAR and the sDPM-MAR approach has the best concordance, risk $R^2$, and PVC.  However, the results from sDPM-MNAR are nearly as good as those from sDPM-MAR and not significantly different for any of the measures.  Thus, very little is lost by including the mising indicators into the imputation strategy.  Both of these approaches are decidedly better than the other approaches insofar as predictive accuracy.  However, the MVN-MAR and MVN-MNAR approaches are not significantly different in terms of variable selection.  Indicating that, for this case at least, the MVN assumption is somewhat robust to departures.

The data are generated by the MVN-MNAR model in simulation Case 3, and unsurprisingly the MVN-MNAR approach has the best concordance, risk $R^2$, and PVC.  However, once again the results from sDPM-MNAR are not significantly different from the best method.  Both MVN-MNAR and sDPM-MNAR are better than the other approaches for all measures.
In Case 4 where the data are generated by sDPM and the missing data is MNAR, sDPM-MNAR is much better than any of the other approaches for predictive accuaracy.  sDPM-MAR is still possibly as good for variable selecton accuracy, but both are far better for variable selection than all other approaches.  GBM and CPH-SW give more reasonable predictive performance than the MAR approaches in this case as well; they both essentially treat missing-ness as a predictor in the regression.

Case 5 used data generated from the empirical distribution of the BPR data with informative variables MAR to get a feel for how well the sDPM approach performs when the modeling assumptions are not exactly true.  In this case, sDPM-MAR is slightly better than the other methods for prediction accuracy, whereas, sDPM-MNAR is the best method when the informative variables are MNAR in Case 6.  Cases 7 and 8 use the same data generating processes as Cases 5 and 6, repectively, only two of the informative varibles are hidden to the model fitting/prediction procedure.  Once again, sDPM-MAR is the best approach for the MAR case, and sDPM-MNAR is better than all other methods in the MNAR setting.  All of the Bayesian imputation methods significantly outperform the other methods with respect to model size and variable selection.  Interestingly, in cases 7 and 8 these approaches want to use an average of 6-7 variables in the model, when there are only four variables available for selection that are truly part of the data generating model.  Clearly the models are trying to replace the two missing predictors with {\em surrogate} variables.  Given that sDPM-MNAR is the only approach that gives nearly identical performance to the best method in all eight simulation cases, it is the best choice for similar missing data regression problems.

\begin{figure}[t!]
\vspace{-.1in}
\centering
\caption{Coefficients on $x_1, x_2, x_3, x_{26}, x_{27}, x_{28}$ in simulation case 3: MVN with informative missingness.  Solid vertical lines are placed at the {\em true} value for each of the respective parameters.  Methods are colored blue if their mean squared error (MSE) for estimation of the respective parameter are not statistically different from the method with the best MSE.}
\vspace{-.04in}
  \includegraphics[width=.92\textwidth, height=.369\textheight]{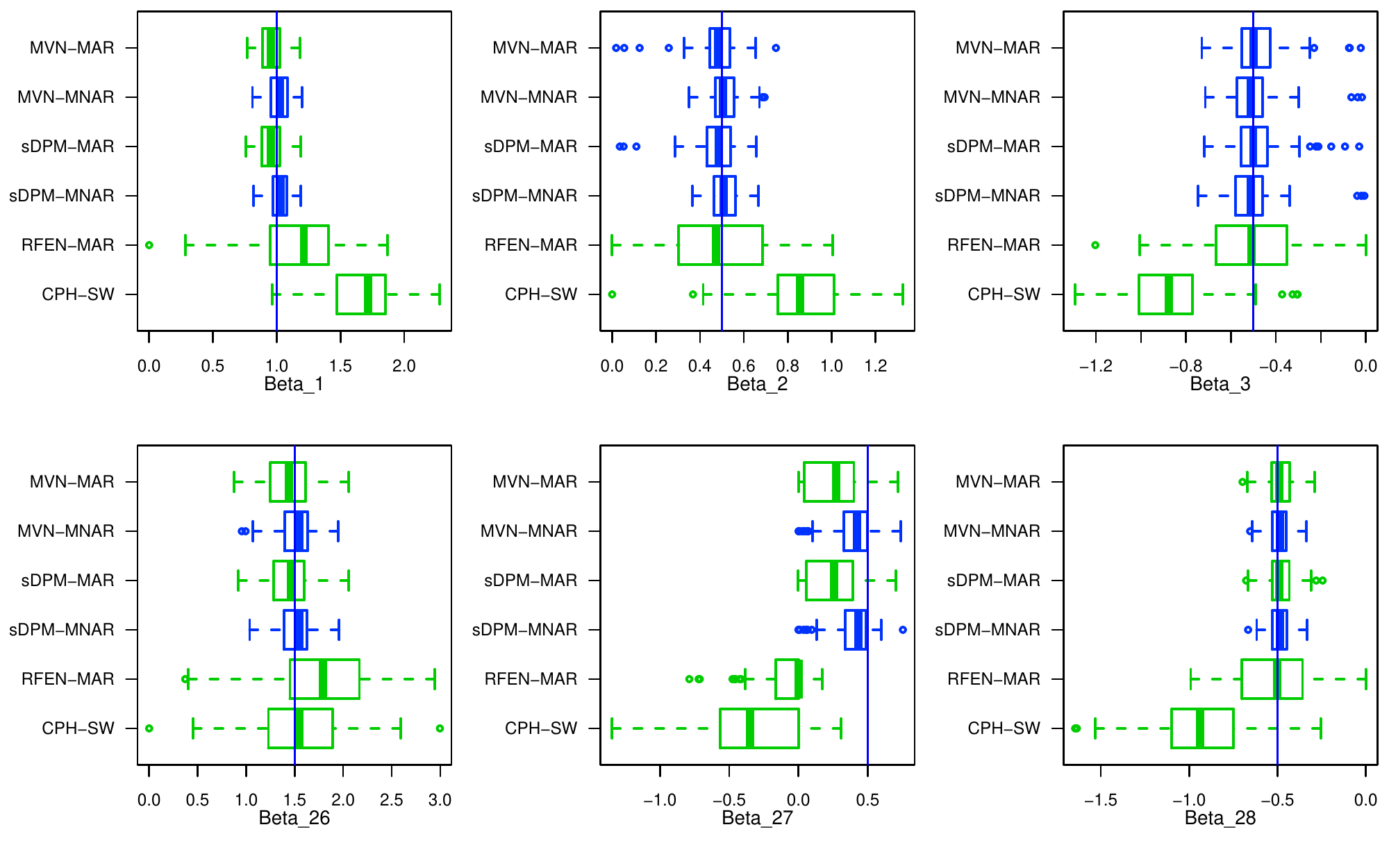}
\label{fig:MNAR_box}
\vspace{-.2in}
\end{figure}

We end this section with a brief discussion about inference in the presence of missing data.  The primary goal of the BPR study is prediction, but in many problems inference is the primary objective.  It is useful to assess the importance of certain variables in a predictive model as well, both for subjective validation and to assist in the acceptance and implementation of the approach.  Figure~\ref{fig:MNAR_box} provides boxplots of the estimated $\beta_j$ coefficients for each method on Case 3: MVN with MNAR observations.  A word of warning is apparent here; do not use a missing indicator / missing factor level in a regression analysis if infernence is one of the goals.  The coefficients obtained with the CPH-SW approach using this strategy are badly biased.  This is because the $\beta_j$ estimate is essentially only influenced by the observed $x_j$; the observations with unobserved $x_j$ serve to estimate the missing-ness indicator coefficient.

Another interesting takeaway is that the MNAR methods do slightly better than their MAR counterparts for estimation of the $\beta_j$, but not by much.  This may seem surprising in light of the results in Table~\ref{tab:sim}(c) and Figure~\ref{fig:risk_Rsq}, showing the MNAR methods to be substantially better at prediction.  This is due to the fact that some of the information about the missing $x_j$ is necessarilly contained in the response $y$, thus making it {\em almost} a MAR situation for the purpose of inference.  Therefore, reasonably accurate estimates of the $\beta_j$ can be obtained for these variables from a MAR approach.  However, when the $y$ variable is also missing (i.e., prediction of a new observation) then the MNAR nature of the $x_j$ is much more of a problem as the imputation of $x_j$ can no longer make use of the dependance on $y$.  The MVN-MNAR and sDPM-MNAR approaches discussed above allow for a dependancy between the indicator of missing-ness variable and the $x_j$ which can be estimated in part from the relationship of $x_j$ with $y$.  Thus, the MNAR approaches make explicit use of the fact that $x_j$ is missing when imputing and ultimately making a prediction of $y$.  This allows for better prediction results than the MAR approaches in MNAR situations, even if the estimates of the $\beta_j$ were the same.

\vspace{-.13in}
\section{Application to Bedside Patient Rescue}
\vspace{-.09in}
\label{sec:BPR}

The BPR data set has 171 predictor variables in total, 75 of which are discrete, and 10 targeted interactions of the variables were included as well.  A glossary of the predictor variables is provided in the Supplementary Material.  There are 34,923 total observations, 3,186 of which had an {\em event} (i.e., cardiac arrest, transfer to the intensive care unit (ICU), or the patient requiring rapid response team intervention).  The most recent 5,000 observations were held out from model fitting to be a validation test set.  The performance of the various models in terms of concordance of predicted risk to the test set event (and censored) times is provided in Table~\ref{tab:BPR_concord} along with estimated standard errors.  All of the linear regression models (i.e., all but GBM) also included square terms for each continuous predictor to allow for quadratic behavior (particularly across lab values) if necessary. For the proposed Bayesian methods, the predicted risk is a distribution, but was collapsed into the posterior mean for this comparison.  The sDPM-MNAR approach had the highest out-of-sample-concordance.  Methods were compared against one another via a bootstrap procedure to generate individual 95\% CIs.  Those comparisons to sDPM-MNAR that had CI's including 0 are bolded, i.e., only MVN-MNAR.  It appears as though the MNAR models are slightly preffered in this case.

\renewcommand{\arraystretch}{1.0} 
\begin{table}
\footnotesize
  \centering
  \vspace{-.1in}
  \caption{Out-of-sample concordance of predicted $\hlambda$ to test data set $y_i$, along with stardard error (SE) estimate and model size $p$.  Methods not significantly different from the {\em best} method according to a comparison via uncorrected 95\% bootstrap CIs are in bold.}
  \label{tab:BPR_concord}
  \vspace{-.1in}
  \begin{tabular}{lcc}
    \hline
    Method &   Concordance (SE)$\;\;\;$ & $p$ \\
    \hline
MVN-MAR    &    0.876  $\:$($\:$0.0092$\:$) & 46.4\\
\bf MVN-MNAR   &    \bf 0.881 (0.0087) & 39.8\\
sDPM-MAR   &    0.875 $\:$($\:$0.0093$\:$) & 51.8\\
\bf sDPM-MNAR  &    \bf 0.885 (0.0089) & 50.5\\
GBM        &    0.874 $\:$($\:$0.0099$\:$) & 84.0\\
RFEN       &    0.877 $\:$($\:$0.0093$\:$) & 44.0\\
CPH-SW     &    0.850 $\:$($\:$0.0104$\:$) & 93.0\\
\hline
  \end{tabular}
  \vspace{-.0in}
\end{table}
\renewcommand{\arraystretch}{.95} 

\begin{figure}[t!]
  \vspace{-.1in}
\centering
\caption{Out-of-sample classification performance. (a) ROC curves for BPR using the predicted hazard rate $\hlambda$ to classify the binary event outcome on the test data set. (b) BSPR predicted scores on the test data set by outcome, with the $\sim$1/10 alarm cut off as a dashed line.}
\vspace{-.11in}
    \begin{subfigure}[b]{.528\textwidth}
      \centering
      \caption{}
\vspace{-.13in}
      \includegraphics[width=.91\textwidth]{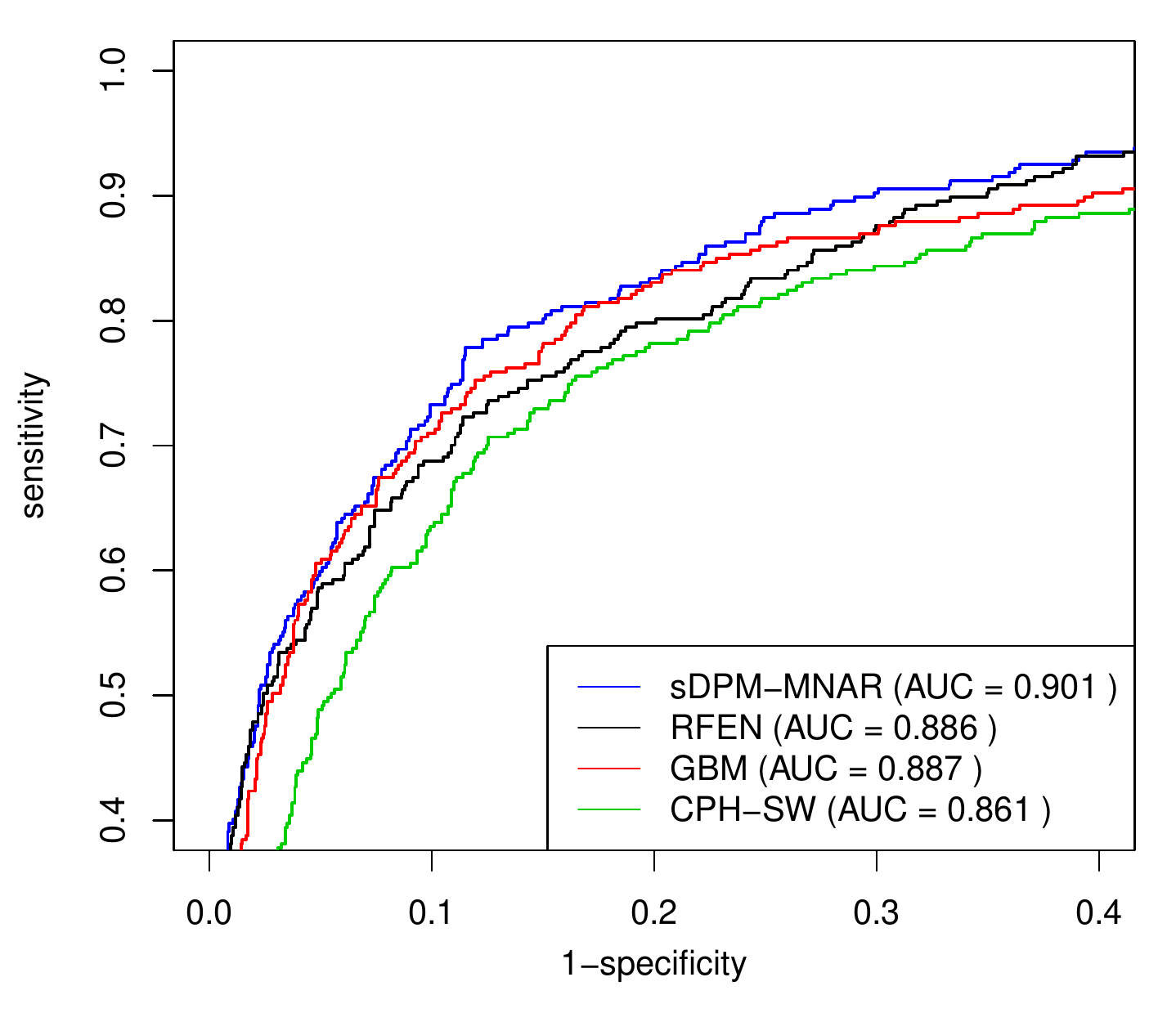}
    \end{subfigure}
    \begin{subfigure}[b]{.459\textwidth}
      \centering
      \caption{}
\vspace{-.13in}
      \includegraphics[width=.91\textwidth]{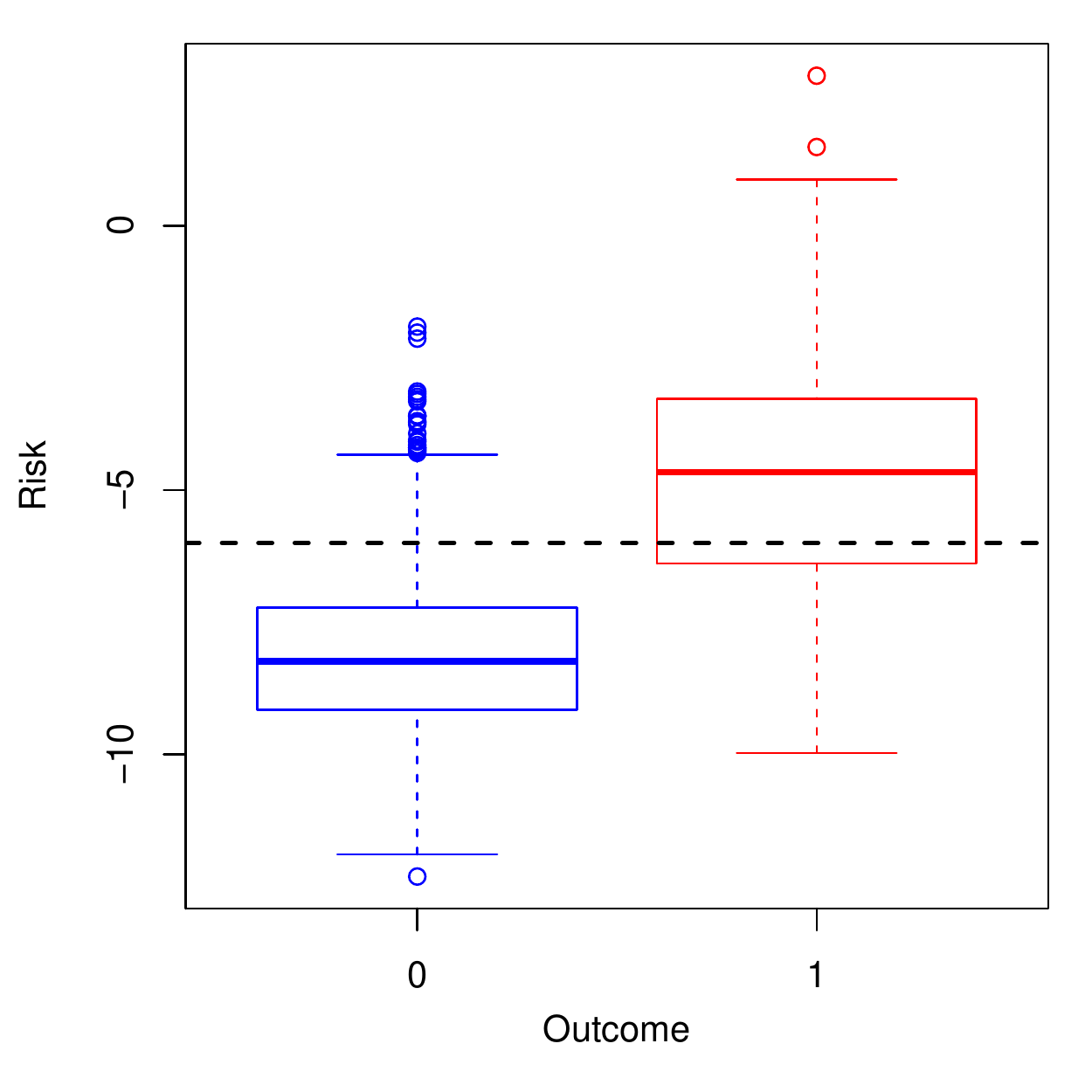}
    \end{subfigure}
\label{fig:class_results}
  \vspace{-.35in}
\end{figure}

The predicted risk scores from the time to event model were also used to classify the binary variable event (or not) on the test data, i.e., ignoring the time to event.  The resulting ROC curves for several methods are displayed in Figure~\ref{fig:class_results}(a). Figure~\ref{fig:class_results}(b) displays boxplots of the risk scores from the sDPM-MNAR method by the event/non-event groups with a dashed threshold line corresponding to a 1/10 false alarm rate.

It is useful to evaluate the importance of the individual predictors on the model, however,
with so many (confounded) predictors this becomes complicated.  We consider two measures of variable importance here, (i) the probabilitly that the variable (or it's square term) were included in the model $\Pr(\delta_j \neq 0)$, and (ii) the main effect index based on the variance decomposition approach to sensitivity analysis \citep{Helton06,Storlie12b}.  Variance decomposition makes use of the law of total variance,
\vspace{-.19in}\bdm
\Var(\lambda(\bx)) = \E \left[ \Var (\lambda(\bx) \mid x_j)\right] + \Var\left[\E (\lambda(\bx) \mid x_j)\right].
\vspace{-.19in}\edm
The main effect index for the $j\th$ predictor is defined as
\vspace{-.15in}\bdm
s_j = \frac{\Var\left[ \E (\lambda(\bx) \mid x_j)\right]}{\Var(\lambda(\bx))}.
\vspace{-.16in}\edm
When $\bx$ consists of independent random variables, $s_j$ can be interpreted as the proportional contribution to the variance in the output ($\lambda(\bx)$ in this case) that can be attributed to $x_j$ alone.
However, when there is correlation among the $\bx$, $s_j$ is not the effect of $x_j$ alone.  It is possible that $\lambda(\bx)$ is not even a function of $x_j$, but $s_j$ is still high because $\lambda(\bx)$ depends heavily on a variable $x_k$ that also has a strong relationship with $x_j$.  The $\Pr(\delta_j \neq 0)$ on the other hand should not be large in such a case, as $x_j$ would not be needed in the model.  Thus, it is beneficial to consider both measures.
Table~\ref{tab:imp_pred} provides estimates of $\Pr(\delta_j \neq 0)$ and $s_j$ for those predictors with $\Pr(\delta_j \neq 0) > 0.5$ and $\hs_j > 0.02$.  The estimates $\hs_j$ are obtained via a scatterplot smoother of the values of $\lambda(\bx)$ across $x_j$, for each posterior realization, and then taking the mean.  The 95\% CI for $s_j$ is obtained by the resulting posterior quantiles.

\begin{table}[t!]
  \vspace{-.11in}
  \centering
 \caption{Marginal posterior inclusion probability for predictors, along with the main effect index $s_j$.  Only predictors with $\Pr(\delta_j \neq 0) >0.5$ and $\hs_j > .02$  are listed.}
 \label{tab:imp_pred}
  \vspace{-.09in}
 {\footnotesize
     \begin{tabular}{lccc}
\hline
Variable & $\Pr(\delta_j \neq 0)$ & $\hs_j$ & $s_j$ 95\% CI \\      
\hline
kirkland.probability & 1.00 & 0.405 & (0.389$\:,\:$0.428)\\
int.rr.spo2.o2flow & 1.00 & 0.370 & (0.354$\:,\:$0.390)\\
max24.int.rr.spo2.ratio & 1.00 & 0.352 & (0.330$\:,\:$0.369)\\
resp.rate & 1.00 & 0.291 & (0.277$\:,\:$0.302)\\
supp.oxygen & 1.00 & 0.264 & (0.245$\:,\:$0.284)\\
range24.int.hr.hemog.ratio & 1.00 & 0.259 & (0.235$\:,\:$0.281)\\
hr & 1.00 & 0.198 & (0.187$\:,\:$0.210)\\
range24.sbp & 1.00 & 0.174 & (0.160$\:,\:$0.188)\\
int.hr.hemog.ratio & 1.00 & 0.155 & (0.142$\:,\:$0.170)\\
lab.val.po2 & 1.00 & 0.128 & (0.106$\:,\:$0.150)\\
sbp & 1.00 & 0.118 & (0.110$\:,\:$0.126)\\
frailty.braden.skin.score & 1.00 & 0.108 & (0.096$\:,\:$0.124)\\
modrass & 1.00 & 0.090 & (0.083$\:,\:$0.097)\\
lab.val.creat & 0.99 & 0.079 & (0.070$\:,\:$0.090)\\
lab.val.troponin & 0.99 & 0.055 & (0.045$\:,\:$0.065)\\
lab.val.bilitot & 0.99 & 0.048 & (0.036$\:,\:$0.060)\\
lab.val.leuko & 0.99 & 0.046 & (0.038$\:,\:$0.054)\\
temp & 0.99 & 0.042 & (0.036$\:,\:$0.048)\\
lab.val.sodium & 0.99 & 0.040 & (0.033$\:,\:$0.047)\\
lab.val.pco2 & 0.98 & 0.038 & (0.032$\:,\:$0.046)\\
iv.sol.drug.dose.4hr & 0.98 & 0.025 & (0.017$\:,\:$0.035)\\
min24.sbp & 0.97 & 0.138 & (0.128$\:,\:$0.154)\\
max24.int.rr.spo2.ratio & 0.96 & 0.296 & (0.273$\:,\:$0.319)\\
map & 0.93 & 0.120 & (0.108$\:,\:$0.130)\\
min24.hr & 0.86 & 0.059 & (0.051$\:,\:$0.069)\\
int.diuret.bun.creat & 0.76 & 0.066 & (0.054$\:,\:$0.079)\\
\hline
     \end{tabular}
     }
  \vspace{-.1in}
\end{table}

\begin{figure}[t!]
\vspace{-.1in}
\centering
\caption{Treating uncertainty in missing predictors.}
\label{fig:get_missing}
\vspace{-.5in}
\centering
\includegraphics[width=.71\textwidth]{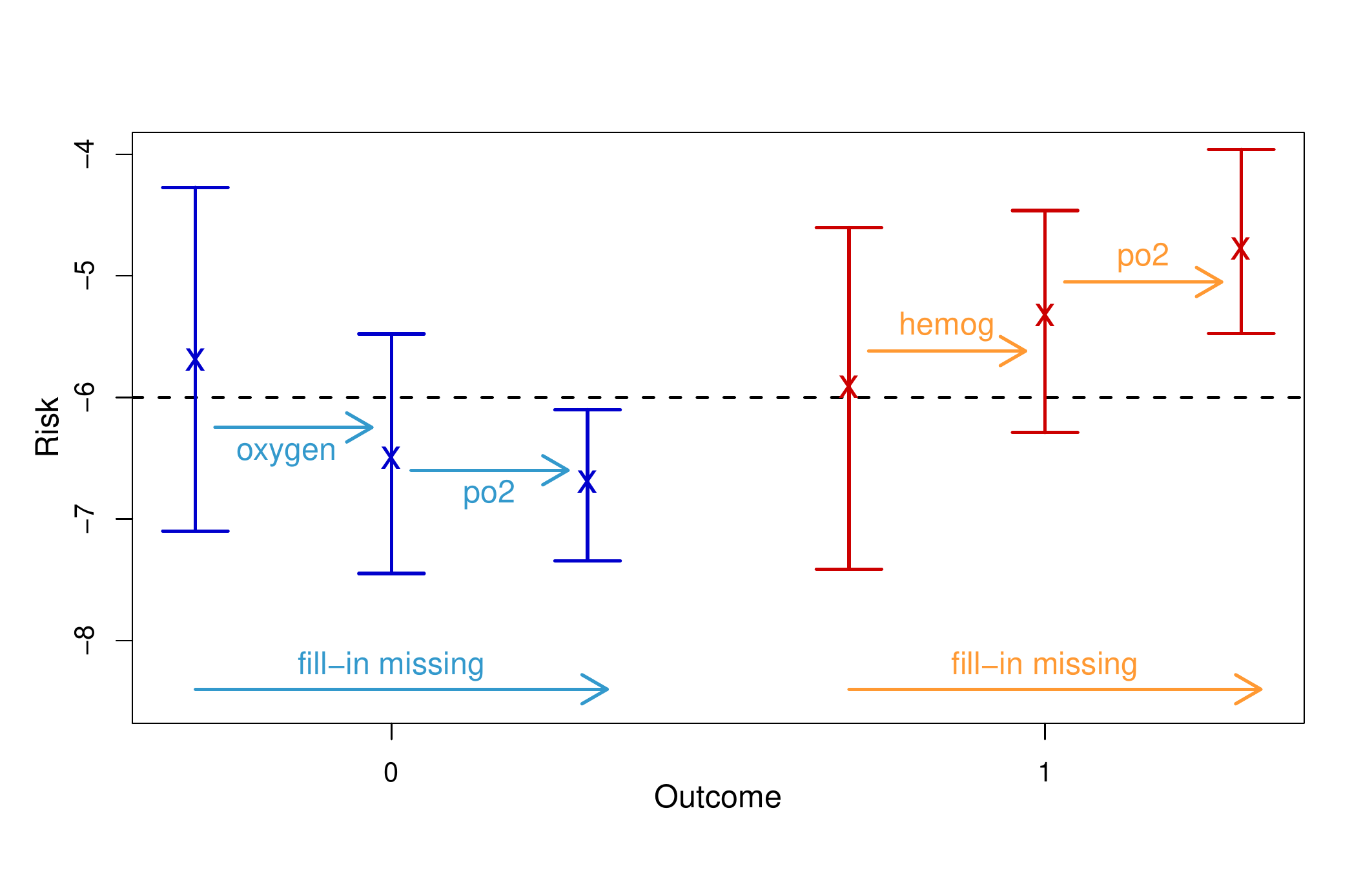}
\vspace{-.27in}
\end{figure}

The variance decomposition concept can also be extended in a very convenient manner to assess the influence of a missing predictor to a given case.  In particular, in the Bayesian framework the risk $\lambda(\bx_{\new})$ for a new observation $\bx_{\new}$ has a posterior predictive distribution, the variability of which can be driven largely by which predictors are missing in $\bx_{\new}$.  The influence of a missing predictor $x_{\new,j}$ on the prediction is defined here as,
\vspace{-.15in}\beq
I_j = \frac{\Var\left[ \E (\lambda(\bx_{\new}) \mid x_{\new,j})\right]}{\Var(\lambda(\bx_{\new}))}.
\label{eq:Ij}
\vspace{-.15in}\eeq
The $I_j$ can be interpreted as the proportion of the variance in $\lambda(\bx_{\new})$ that is due to a lack of knowledge of the value of $x_{\new,j}$, i.e., the variance of $\lambda(\bx_{\new})$ can be reduced by this proportional amount if one were able to obtain the missing value of $x_{\new,j}$.  Thus, $I_j$ can be used to prioritize further collection of obtainable missing predictors for borderline cases.  This concept is illustrated in Figure~\ref{fig:get_missing} where the predictive distribution is provided for a patient that had an event and a patient with a similar risk score that did not have an event.  Missing predictors were ranked according to $I_j$.  In a greedy fashion, the missing predictor with the largest $I_j$ was ``obtained'' by generating a random value from the conditional distribution.  The remaining missing predictors were then assessed again on the basis of $I_j$ until the credible interval did not include the -6 threshold value.  Thus, the original credible intervals are from real patient observations, but the actual missing values could not really be obtained from the retrospective data.  However, these influential lab values and other missing variables {\em could} be collected for future cases, and the contribution to the variance provides a means to target this effort.

\vspace{-.25in}
\section{Conclusions \& Further Work}
\vspace{-.09in}
\label{sec:conclusions}

A principled approach to regression analysis with missing data is proposed within a non-parametric Bayesian framework.  A sparse DPM was used to model the joint distribution of mixed discrete/continuous predictors $\bx$ in a flexible manner.  The Bayesian approach has a lot to offer for missing data problems, e.g., a probabilistically proper treatment and quantification of uncertainty.  Analysis of the BPR problem and a simulation study set up to resemble the BPR problem demonstrated markedly improved inference and prediction capability for the proposed approach over traditional methods.  Perhaps most importantly, the proposed approach allows for the attribution of uncertainty in missing data for a given subject to a given predictor.  This allows for a convenient means of targeting potential data collection efforts to decrease uncertainty.  This was illustrated in the context of the BPR problem, where missing values were ``obtained'' in a greedy fashion.  However, much could be gained from a formal Bayesian decision framework \citep{Berger2013} to the collection of missing information, e.g., in some cases it may be more efficient to gather multiple missing values at one time.

One limitation to the proposed approach is that the MVN and Gaussian mixtures do not scale particularly well to very high dimensional $\bx$ (i.e., $p>300$).  However, \cite{Talhouk2012} discuss the use of a Gaussian graphical model \citep{Giudici1999,Wong2003} for the covariance in the context of the multivariate probit model.  It may be possible to leverage this framework with the sparse DPM approach discussed here to allow for more efficient computation in higher dimensions.

{\small
  \singlespacing
  \vspace{-.23in}
\bibliography{curt_ref.bib}
\bibliographystyle{rss}
}

\newpage

\begin{appendix}

\setcounter{equation}{0}
\setcounter{page}{1}
\renewcommand{\theequation}{A\arabic{equation}}

\vspace{-.2in}
\begin{Large}
\noindent
{\bf
Supplementary Material: ``Prediction and Inference with Missing Data in Patient Alert Systems''
}
\end{Large}
\vspace{-.3in}

\section{MCMC Algorithm and Full Conditionals}
\vspace{-.05in}
\label{sec:computation_supp}

This section describes the MCMC sampling scheme for the full model described in Section~\ref{sec:model_descr} of the main paper.
The complete list of parameters in the model described in (\ref{eq:reg_model}) and (\ref{eq:DPM_model}) are
\vspace{-.05in}\beq
\Theta = \left\{
\btheta,
\left\{\pi_{h}, \bmu_{h}, \bQ_{h}\right\}_{h=1}^H,
\varphi, \eta, \psi, 
\bX^*,
\by^*,
\bphi,
\bgamma,
\right\},
\label{eq:param_list_2}
\vspace{-.05in}\eeq
where $\btheta$ are the regression model parameters ($\btheta=\{\bbeta,\tau^2,\kappa\}$ for this application), $\bX^* = \{\bx_1^*,\dots,\bx_n^* \}'$ and $\bx_i^*$ are the values of latent vector $\bx^*$ for the $i\th$ observation. The $\by^*$ is a vector of the uncensored $y_i$ observations, i.e., $y_i^*=y_i$ for those observations with an event, and $y_i^* \geq y_i$ for right censored observations.  The inclusion of $\by^*$ facilitates the sampling of the $\bbeta$ vector described below.
The posterior distribution of these parameters is approximated via Markov chain Monte Carlo (MCMC).  This is facilitated by including a further latent variable $\bphi = [\phi_1,\dots,\phi_n]'$ to denote which mixture component produced each of the $\bx^*_i$.  For convenience of computation, the number of components in stick-breaking model for $\pi_{h}$ is capped at a finite value $H$, i.e., $h=1,\dots,H$.  The value of $\pi_{h}$ is observed and $H$ can be increased if needed such that approximately half of the $\pi_{h}$ values were negligible ($H\approx 30$ for the BPR problem considered here).

With the inclusion of the $\bphi$, conditional on the {\em rest}, $\pi_h$ has a conjugate update.  The $\bgamma$ must be sampled together with the $\tbmu_h$ and $\tbQ_h$ because their dimensionality changes with the value of $\bgamma$, which requires a reversible jump MCMC update \citep{Green1995,Green2009}.  This is streamlined by the fact that there is a conjugate update for the $\tbmu_h$ and $\tbQ_h$ conditional on $\bgamma$, thus providing a convenient means of producing reasonable proposals for the the $\tbmu_h$ and $\tbQ_h$, independent of the current state; this makes the Jacobian for the dimension matching transformation trivially equal to one.  As stated previously, the likelihood identified parameters $\bmu$ and $\bQ$ are simply computed via deterministic relations after the update of $\tbmu_h$ and $\tbQ_h$.  The latent indicator $\bphi$ has a convenient form for a full conditional to facilitate Gibbs updates.  The scalar parameters $\varphi$, $\eta$, and $\psi$ do not have convenient full conditional forms and are thus updated with a typical random walk MH on log scale.

The updating of $\btheta$ is regression model dependent.  However, whatever the typical update is for the given model in the complete data case would apply, since the current value for $X_{\miss}$ is treated as given.  For example, if the regression model is a normal error, linear model, then the $\btheta$ will have a usual conjugate normal (and inverse gamma) updates.  In the case of the regression model in (\ref{eq:weibull_reg}) there is no such conjugate update.  However, an efficient MH update without the need for a tuning parameter is described below.

The $\kappa$ parameter is updated with a typical random walk MH on log scale and thus $\kappa$, $\varphi$, $\eta$, and $\psi$ are the only updates for the BPR application that require specification of a tuning parameter.  However, these parameters are all scalars so this is fairly straight-forward.

Full conditional distributions with which to perform the Gibbs updates are provided below for some of the parameters listed in (\ref{eq:param_list_2}).  For all other parameters, the specifics of the MH update is described instead.\\[-.1in]

\noindent
{$\underline{\ty_i \mid \mbox{rest}}$}

\noindent
Generate a set of uncensored observations $\ty_i$, $i=1,\dots,n$.
Set $\ty_i=y_i$ for all non-censored $y_i$.
For those $y_i$ which are right-censored, simply compute $\lambda(\bx_i)$ from (\ref{eq:weibull_reg2}), then generate $\ty_i$ as
\vspace{-.1in}\begin{eqnarray}
  U & \sim & \mbox{Unif}(0,1) \nonumber \\
  z & = & 1-U \left[1 - F(y_i; \lambda(\bx_i), \kappa)\right] \nonumber \\
  \ty_i & = & F_i^{-1}(z),\nonumber\\[-.4in] \nonumber 
\end{eqnarray}
where $F(\cdot,\lambda,\kappa)$ is the CDF of a Weibull distribution.\\[-.1in]

\noindent
{\underline{MH update for $\bbeta$}}

\noindent
The coefficients $\bbeta_j= [\beta_{1,1},\beta_{1,2}, \dots, \beta_{j,L_j-1}]'$ are block updated for each $j=1,\dots,p+K$.
Transform the response
\bdm
\ty_i^{(j)}= -\log(y_i^*) - \sum_{k \neq j} \sum_{l=1}^{L_j-1}\beta_{k,l} z_{i,k,l} + \frac{1}{\kappa} \digamma(1) \Gamma(1),
\edm
where $z_{i,j,1}=x_{i,j}$ and $L_j \stackrel{\mbox{\scriptsize def}}{=} 2$ for $j>r$ and $z_{i,j,l}=I_{\{x_{i,j}=l\}}$ for $j<r$.
Then,
\vspace{-.1in}\begin{eqnarray}
  \E(\ty_i^{(j)}) & = & \sum_{l=1}^{L_j-1}\beta_{j,l} z_{i,j,l}  \nonumber \\
  {\tsigma}^2 = \Var(\ty_i^{(j)}) & = & \frac{1}{\kappa^2} \left\{\Gamma(1) \left[\trigamma(1)+\digamma^2(1) \right] - \digamma^2(1) \Gamma^2(1) \right\},\nonumber\\[-.4in] \nonumber 
\end{eqnarray}
where $\Gamma$, $\digamma$ and $\trigamma$ are the gamma, digamma and trigamma functions, respectively.
Thus an effective strategy to update $\bbeta_j$ is to draw a $\delta_j$ conditional on the current $\bbeta_j$ and if $\delta_j=1$, then draw a proposal $\bbeta_j$ from the conjugate normal update assuming $\ty_i^{(j)}$ are normally distributed with mean and variance above.  Finally, accept or reject the proposed $\bbeta_j$ in an MH step under the Weibull likelihood.  This approach requires no tuning and resulted in high acceptance rates for all $\bbeta_j$ (lowest acceptance rate of 55\% in the BPR analysis with $\sim$200 predictors).

Specifically, to generate a proposal $\bbeta_j^{\pr}$, first draw $\delta_j^{\pr} \sim \mbox{Bernoulli}(p_{01}I_{\{\delta_j=0\}}+p_{11}I_{\{\delta_j=1\}})$;  we set $p_{01}=0.3$ and $p_{11}=0.7$ for all results in the paper.  If $\delta_j^{\pr} =0$, then set $\bbeta_j^{\pr}=\bzero$.  Otherwise, draw $\bbeta_j^{\pr} \sim \cN(\bmu_j^{\pr}, \bSigma_j^{\pr})$, where
\vspace{-.1in}\begin{eqnarray}
\bSigma^{\pr}_j & = & {\tsigma}^{2} \left[\bX_j'\bX_j + \left(\frac{\tsigma}{\tau}\right)^2\bI \right]^{-1}\nonumber \\ 
\bmu^{\pr}_j & = & {\tsigma}^{-2} \bSigma^{\pr} \bX_j \tby^{(j)}. \nonumber \\[-.4in] \nonumber 
\end{eqnarray}
Let $d(\bbeta_j^{\pr} \mid \bbeta_j)$ represent the density of this proposal.  The MH ratio is then
\vspace{-.15in}\bdm
MH = \frac{f(\by \mid \bbeta^{\pr}, \bX^*, \kappa) f(\bbeta_j^{\pr}) d(\bbeta_j \mid \bbeta_j^{\pr})}
{f(\by \mid \bbeta, \bX^*, \kappa) f(\bbeta_j) d(\bbeta_j^{\pr} \mid \bbeta_j)},
\edm
where $f(\by \mid \bbeta, \bX^*, \kappa)$ is the marginal likelihood for the regression model in (\ref{eq:weibull_reg}) and $f(\bbeta)$ is the density of the prior distribution for $\bbeta$, i.e., that provided in (\ref{eq:beta_prior}).  In the results of the main paper, the prior probability of inclusion $\rho_j$ was set to 0.5 for all $j$.\\[-.1in]

\noindent
{$\underline{\tau^2 \mid \mbox{rest}}$}

\noindent
Conditional on the rest, $\tau^2$ has a simple conjugate Inverse-Gamma ($\cI\cG$) update,
\bdm
\tau^2 \sim \cI\cG\left(A_\tau + \frac{J}{2}\;,\; B_\tau + \frac{1}{2}\sum_{j=1}^p \sum_{k=1}^{L_j-1} \beta_{j,k}^2\right),
\edm
where $L_j$ is the number of levels for the categorical $x_j$ and $L_j\stackrel{\mbox{\scriptsize def}}{=} 2$ for continuous $x_j$ (i.e., there is only one $\beta_{j,1}$ to sum over in the above expression), and $J = \sum_{j=1}^p (L_j-1)$.
\\[-.1in]

\noindent
{\underline{MH update for $\kappa$}}

The $\kappa$ parameter is updated via a MH random walk on log scale, i.e., $\log(\kappa^{\pr}) = \log(\kappa+\epsilon)$ for a deviate $\epsilon \sim \cN(0,s^2)$. A tuning parameter $s=0.05$ was used to achieve an acceptance rate of 40\%.  Let the density of the proposal, given the current value of $\kappa$ be denoted $d({\kappa}^{\pr} \mid {\kappa})$.  The only portion of the full model likelihood that differs between the current value and the proposal is $f(\by \mid \bbeta, \bX^*, \kappa)$.
The MH ratio is then
\bdm
MH = \frac{f(\by \mid \bbeta, \bX^*, \kappa^{\pr}) f(\kappa^{\pr}) d(\kappa \mid \kappa^{\pr})}
{f(\by \mid \bbeta, \bX^*, \kappa) f(\kappa) d(\kappa^{\pr} \mid \kappa)},
\edm
where $f(\kappa)$ is the density of the prior distribution for $\kappa$, i.e., Gamma$(A_\kappa, B_\kappa)$.
\\[-.1in]

\noindent
{$\underline{\bpi \mid \mbox{rest}}$}

\noindent
There is a one to one correspondence between $\bpi$ and $\bv = [v_{1},\dots,v_{H}]'$ in (\ref{eq:v_m}).  Conditional on the rest of the parameters, $\bv_j$ depends only on $\bphi$ and $\varpi$.  Specifically,
\bdm
v_{h} \mid \mbox{rest} \stackrel{ind}{\sim} \mbox{Beta}(a^*_h, b^*_h),
\edm
where,
\vspace{-.1in}\begin{eqnarray}
a^*_h & \!=\!& \sum_{i=1}^{n}  I_{\{\phi_i=h\}} +1 \nonumber \\
b^*_h & \!= \!& \sum_{i=1}^{n} I_{\{\phi_i>h\}} + \varpi \nonumber \\[-.4in] \nonumber
\end{eqnarray}
\\[-.1in]

\noindent
{\underline{MH update for $\bgamma, \bmu_{h}, \bQ_{h}$}}

\noindent
A proposal for the $\bgamma$ vector is obtained via an add, delete, or swap move. That is, the proposal $\bgamma^{\pr}$ is generated as follows.
\begin{itemize}
\item[(i)] Set the proposal $\bgamma^{\pr}=\bgamma$
\item[(ii)] Randomly choose an integer $j^{\pr}$ from $1,\dots,p$.
\item[(iii)] Flip the value of $\gamma_{j^{\pr}}$, i.e., $\gamma_{j^{\pr}}^{\pr}=1-\gamma_{j^{\pr}}$.
\item[(iv)] If the set $\{j:\gamma_j \neq \gamma_{j^{\pr}}\}$ is not empty, draw a Bernoulli $B^{\pr}$ with probability $\pi$.
\item[(v)] If $B^{\pr}=1$ randomly choose another $j^{**}$ from the set $\{j:\gamma_j \neq \gamma_{j^{\pr}}\}$ and also set $\gamma^{\pr}_{j^{**}}=1-\gamma_{j^{**}}$, i.e., a swap proposal.  If $B^{\pr}=0$, leave $\bgamma^{\pr}$ as a single variable add/delete proposal.
\end{itemize}
Let $d(\bgamma^{\pr} \mid \bgamma)$ represent the density of this proposal.

Now a proposal for $\vartheta^{\pr} = \{\bmu^{\pr}_{h}, \bQ_{h}^{\pr}\}_{h=1}^H$ conditional on the proposed $\bgamma^{\pr}$ is drawn in the following manner.  Conditional on the rest of the parameters and data, $\bmu_{h}, \bQ_{h}$, $h=1,\dots,H$ depend only on $(\bgamma,\bphi, \bX^*, \varphi, \eta, \psi)$.  In the Supplemental Material of \cite{Storlie17b} it was shown that
\vspace{-.1in}\begin{eqnarray}
\bSigma_{h11}  \mid \mbox{rest}  & \!\sim\!& \cI\cW(n_h+\eta-p_2 \:,\: \bV_{h11})\nonumber \\
\bmu_h \mid \mbox{rest}  & \!\sim\!& \cN\left(\frac{n_h}{n_h\!+\!\varphi}\bar{\bx}_{h1}\:,\: \frac{1}{n_h+\varphi}\bSigma_{h11} \right), \nonumber \\[-.4in] \nonumber
\end{eqnarray}
where $n_h=\sum I_{\{phi_i=h \}}$, $\bx^{(1)}_i = \{x^*_{i,j}: \gamma_j=1\}$, $p_2=\sum_j I_{\{\gamma_j=1\}}$, and
\vspace{-.1in}\begin{eqnarray}
\bar{\bx}_{h1} &\!=\!& \frac{1}{n_h}\sum_{\phi_i=h} \bx^{(1)}_{i}\nonumber \\
\bV_{h11} &\!=\!& \sum_{\phi_i =h} (\bx^{(1)}_{i} - \bar{\bx}_{h1})(\bx^{(1)}_{i} - \bar{\bx}_{h1})' + \frac{n_h\varphi}{n_h\!+\!\varphi}\bar{\bx}_{h1}\bar{\bx}_{h1}' + \bPsi_{11}.\nonumber \\[-.4in] \nonumber
\end{eqnarray}
Also,
\vspace{-.1in}\begin{eqnarray}
\bQ_{22}  \mid \mbox{rest}  & \!\sim\!& \cW(n+\eta \:,\: \bV_{\!22 \mid 1}) \nonumber \\
\bQ_{21} \mid \mbox{rest}  & \sim & \cM\cN(-\bQ_{22}\bV_{\!\!21} \bV_{\!11}^{-1} \:,\: \bQ_{22} \:,\: \bV_{\!\!11}^{-1}) \nonumber \\
\bb_2 \mid \mbox{rest}  & \!\sim\!& \cN\left(\frac{n}{n\!+\!\varphi}(\bQ_{22} \bar{\by}_2 +  \bQ_{21} \bar{y}_1) \;,\; \frac{1}{n\!+\!\varphi}\bQ_{22}  \right), \nonumber \\[-.4in] \nonumber
\end{eqnarray}
where $\bx^{(2)}_i = \{x^*_{i,j}: \gamma_j=0\}$, and
\vspace{-.15in}\begin{eqnarray}
  \bar{\bx}_{1} &\!=\!& \frac{1}{n}\sum_{i=1}^n \bx^{(1)}_{i}, \nonumber \\
  \bar{\bx}_{2} &\!=\!& \frac{1}{n}\sum_{i=1}^n \bx^{(2)}_{i}, \nonumber \\
    \bV_{11} & \! = \! & \sum_{i=1}^n (\bx^{(1)}_{i} - \bar{\bx}_{1})(\bx^{(1)}_{i} - \bar{\bx}_{1})' + \frac{n\varphi}{n\!+\!\varphi}\bar{\bx}_{1}\bar{\bx}_{1}' + \bPsi_{11}, \nonumber \\
    \bV_{22} & \! = \! & \sum_{i=1}^n (\bx^{(2)}_{i} - \bar{\bx}_{2})(\bx^{(2)}_{i} - \bar{\bx}_{2})' + \frac{n\varphi}{n\!+\!\varphi}\bar{\bx}_{2}\bar{\bx}_{2}' + \bPsi_{22}, \nonumber \\
    \bV_{21} & \! = \! & \sum_{i=1}^n (\bx^{(2)}_{i} - \bar{\bx}_{2})(\bx^{(1)}_{i} - \bar{\bx}_{1})' + \frac{n\varphi}{n\!+\!\varphi}\bar{\bx}_{2}\bar{\bx}_{1}' + \bPsi_{21}, \nonumber \\
    \bV_{2 \mid 1} & \! = \! & \bV_{22} - \bV_{21}\bV_{11}^{-1}\bV_{21}'. \nonumber\\[-.475in]
\end{eqnarray}

Thus, draw a proposal $\vartheta^{\pr}$ according to the conjugate update above with $\bgamma = \bgamma^{\pr}$.  Let this proposal distribution be denoted $d(\vartheta^{\pr} \mid \bgamma^{\pr}, \bphi, \bX^*, \varphi, \eta, \psi)$.  Thus in the context of a reversible jump transition, the current set of parameters is $\vartheta=\{\bmu_{h}, \bQ_{h}\}_{h=1}^H$, while the dimension matching random vector for the proposal is $u = \vartheta^{\pr}$, with the mapping $(u^{\pr},\vartheta^{\pr}) = h(\vartheta,u) = (u,\vartheta)$, i.e., the transformation function $h$ is simply a reordering of the vector.  Thus, the Jacobian of the transformation from $(\vartheta,u)$ to $(\vartheta^{\pr},u^{\pr})$ is trivially equal to one.

The reversible jump acceptance probability is then the minimum of 1 and,
\vspace{-.15in}\bdm
MH = \frac{\!\!\!\!\!f(\bX^* \mid \bgamma^{\pr}, \vartheta^{\pr}, \bphi) f(\bgamma^{\pr}) f(\vartheta^{\pr} \mid \bgamma^{\pr}, \varphi, \eta, \psi) \:d(\bgamma \mid \bgamma^{\pr}) \:d(\vartheta \mid \bgamma, \bphi, \bX^*, \varphi, \eta, \psi)\;}
{\:f(\bX^* \mid \bgamma, \vartheta, \bphi) \;\;\;f(\bgamma) \; f(\vartheta \mid \bgamma, \varphi, \eta, \psi) \;\;\; d(\bgamma^{\pr} \mid \bgamma) \:d(\vartheta^{\pr}, \mid \bgamma^{\pr}, \bphi, \bX^*, \varphi, \eta, \psi)}
\edm
where $f(\bX^* \mid \bgamma, \vartheta, \bphi)$ is a product of multivariate normal likelihoods and $f(\bgamma)$ is the prior distribution for $\bgamma$, i.e., independent Bernoulli$(\varrho)$, and $f(\vartheta \mid \bgamma, \varphi, \eta, \psi)$ is the density of the prior distribution defined in (\ref{eq:comp_prior_1}) and (\ref{eq:comp_prior_2}).  In the results of the main paper, $\varrho$ was set to 0.5.
\\[-.1in]

\noindent
{\underline{MH update for $\varphi$}}

\noindent
The $\varphi$ parameter is updated via a MH random walk on log scale, i.e., $\log(\varphi^{\pr}) = \log(\varphi+\epsilon)$ for a deviate $\epsilon \sim \cN(0,s^2)$. A tuning parameter $s=0.2$ was used to achieve an acceptance rate of 40\%.  Let the density of the proposal, given the current value of $\varphi$ be denoted $d({\varphi}^{\pr} \mid {\varphi})$.  The only portion of the full model likelihood that differs between the current value and the proposal is $\prod_{h=1}^H f(\bmu_{h1} \mid \bSigma_{h11}, \varphi) f(\bb_2 \mid \bQ_{22}, \varphi)$.
The MH ratio is then
\bdm
MH = \frac{\prod_{h=1}^H f(\bmu_{h1} \mid \bQ_{h11}, \varphi^{\pr}) f(\bb_2 \mid \bQ_{22}, \varphi^{\pr}) f(\varphi^{\pr}) d(\varphi \mid \varphi^{\pr})}
{\prod_{h=1}^H f(\bmu_{h1} \mid \bQ_{h11}, \varphi) f(\bb_2 \mid \bQ_{22}, \varphi) f(\varphi) d(\varphi^{\pr} \mid \varphi)},
\edm
where $f(\varphi)$ is the density of the prior distribution for $\varphi$, i.e., Gamma$(A_\varphi, B_\varphi)$.
\\[-.1in]

\noindent
{\underline{MH update for $\eta$}}

\noindent
The $\eta$ parameter is also updated via a MH random walk on log scale, i.e., $\log(\eta^{\pr}) = \log(\eta+\epsilon)$ for a deviate $\epsilon \sim \cN(0,s^2)$. A tuning parameter $s=0.5$ was used to achieve an acceptance rate of 40\%.  Let the density of the proposal, given the current value of $\eta$ be denoted $d({\eta}^{\pr} \mid {\eta})$.  The only portion of the full model likelihood that differs between the current value and the proposal is $\prod_{h=1}^H f(\bSigma_{h11} \mid \eta, \psi) f(\bQ_{22} \mid \eta, \psi)$.
The MH ratio is then
\bdm
MH = \frac{\prod_{h=1}^H f(\bSigma_{h11} \mid \eta^{\pr}, \psi) f(\bQ_{22} \mid \eta^{\pr}, \psi) f(\eta^{\pr}) d(\eta \mid \eta^{\pr})}
  {\prod_{h=1}^H f(\bSigma_{h11} \mid \eta, \psi) f(\bQ_{22} \mid \eta, \psi) f(\eta) d(\eta^{\pr} \mid \eta)},
\edm
where $f(\eta)$ is the density of the prior distribution for $\eta$, i.e., Gamma$(A_\eta, B_\eta)$.
\\[-.1in]

\noindent
{$\underline{\psi \mid \mbox{rest}}$}

\noindent
The $\psi$ parameter is also updated via a MH random walk on log scale, i.e., $\log(\psi^{\pr}) = \log(\psi+\epsilon)$ for a deviate $\epsilon \sim \cN(0,s^2)$. A tuning parameter $s=0.5$ was used to achieve an acceptance rate of 40\%.  Let the density of the proposal, given the current value of $\psi$ be denoted $d({\psi}^{\pr} \mid {\psi})$.  The only portion of the full model likelihood that differs between the current value and the proposal is $\prod_{h=1}^H f(\bSigma_{h11} \mid \eta, \psi) f(\bQ_{22} \mid \eta, \psi)$.
The MH ratio is then
\bdm
MH = \frac{\prod_{h=1}^H f(\bSigma_{h11} \mid \eta, \psi^{\pr}) f(\bQ_{22} \mid \eta, \psi^{\pr}) f(\psi^{\pr}) d(\psi \mid \psi^{\pr})}
  {\prod_{h=1}^H f(\bSigma_{h11} \mid \eta, \psi) f(\bQ_{22} \mid \eta, \psi) f(\psi) d(\psi^{\pr} \mid \psi)},
\edm
where $f(\psi)$ is the density of the prior distribution for $\psi$, i.e., Gamma$(A_\psi, B_\psi)$.
\\[-.1in]

\noindent
{\underline{MH update for $\bX^*$}}
\\[-.1in]

The rows of $\bX^*$, $\bx^*_i$, $i=1,\dots,n$ can be updated independently and in parallel.
Recall the definition of $\bx^*=[\bw_1',\dots,\bw_p']'$ above (\ref{eq:w_distn}), and denote $\bx_i^*=[\bw_{i,1}',\dots,\bw_{i,p}']'$.  If $x_{i,j}$ is {\bf not} missing, then the corresponding $\bw_{i,j}$ have simple truncated normal updates for $j \leq r$, and if $j>r$ then $\bw_{i,j} \equiv x_{i,j}$.  However, when $x_{i,j}$ is missing, there is no convenient distributional form for updating the $\bw_{i,j}$, due to the dependence on the response $y_i$.  If, on the other hand, there were no $y_i$ included in the {\em rest}, then $\bw_{i,j} \mid \mbox{rest}$ would have normal updates, conditional on the remaining $\bw_{i,j'}$, $j' \neq j$.  Thus, for updating the $\bw_{i,j}$ corresponding to a missing $x_{i,j}$ an effective strategy is to draw a proposal from the closed form update as if there were no $y_i$ and use this draw as a proposal in a MH step.  This approach requires no tuning and resulted in high acceptance rates for all $x^*_{i,j}$ (mean acceptance was 76\%, lowest was 46\%) in the BPR analysis and similarly high acceptance in all of our simulation study cases.

Specifically, update each $x^*_{i,j}$ as follows,
\begin{itemize}
\item[(i)] If $j \leq r$ and $x_{i,j}$ is not missing, then update each component of $\bw_{i,j}$ from its full conditional distribution by drawing from the normal distribution defined by $\bmu_{\phi_i}$, and $\bQ_{\phi_i}$, conditional on the remaining elements of $\bx^*_i$, such that $w_{i,j,k} < w_{i,j,x_{i,j}}$, for all $k \neq x_{i,j}$, where $w_{i,j,k}$ is the $k\th$ element of $\bw_{i,j}$ and $w_{i,j,0} \equiv 0$.
\item[(ii)] If $j > r$ and $x_{i,j}$ is not missing, then $\bw_{i,j} \equiv x_{i,j}$.
\item[(iii)] If $x_{i,j}$ {\em is} missing, then draw a proposal $\bw_{i,j}^{\pr}$ for each component of $\bw_{i,j}$ from the normal distribution defined by $\bmu_{\phi_i}$, and $\bQ_{\phi_i}$, conditional on the remaining elements of $\bx^*_i$, with no restriction on the lower/upper bound for the $w_{i,j,k}$.  Replace the corresponding elements of $\bx^*_i$ with $\bw_{i,j}^{\pr}$ and denote this vector ${\bx^*_i}^{\pr}$. The $\bw_{i,j}^{\pr}$ is accepted/rejected according to the MH ratio,
\vspace{-.05in}\bdm
MH = \frac{f(y_i \mid \bbeta, {\bx^*_i}^{\pr}, \kappa)}
{f(y_i \mid \bbeta, {\bx^*_i}, \kappa)}
\edm
as the contribution to the posterior for $\bw_{i,j}^{\pr}$ conditional on the remaining elements of $\bx^*_i$ is identical to the proposal density.  Thus, they cancel out and only the ratio of the likelihoods remains.\\[-.1in]
\end{itemize}

\noindent
{$\underline{\bphi \mid \mbox{rest}}$}

\noindent
$\Pr(\phi_i = h \mid \mbox{rest} ) \propto \pi_h \:\cN\left( \bx^*_i; \bmu_h, \bQ_h^{-1} \right)$,
where $\cN( \cdot; \bmu, \bSigma)$ is the normal density with mean $\bmu$ and covariance $\bSigma$.
\\[-.1in]

\section{BPR Variable Descriptions}
\vspace{-.05in}
\label{sec:computation}

\begin{description}
  \small
  \singlespacing
\setlength\itemsep{-.04in}
\item[vitals.height:] Patient height.
\item[vitals.modrass:] RASS mental status score.
\item[frailty.braden.skin.score:] Braden skin score.
\item[frailty.fall.risk.score:] Score indicating patient risk of falling.
\item[frailty.patient.mobility:] Mobility sub-score of the Braden score.
\item[frailty.patient.nutrition:] Nutrition sub-score of the Braden score.
\item[frailty.pt.activity.level:] Activity sub-score of the Braden score.
\item[frailty.sensory.perception:] Perception sub-score of the Braden score.
\item[frailty.getup:] Get-up-and-go test of patient mobility.
\item[lab.value.glucose:] Glucose lab value.
\item[lab.value.bun:] Blood urea nitrogen lab value.
\item[lab.value.creat:] Creatinin lab value.
\item[lab.value.hc03:] Bicarbonate lab value.
\item[lab.value.hemog:] Homoglobin lab value.
\item[lab.value.leuko:] Leukocytes lab value.
\item[lab.value.nphils:] Neutrophils lab value.
\item[lab.value.platelet:] Platelet lab value.
\item[lab.value.potas:] Potasssium lab value.
\item[lab.value.sodium:] Sodium lab value.
\item[lab.value.troponin:] Tropinin lab value.
\item[lab.value.aniongap:] Anion gap lab value.
\item[lab.value.alk:] Alkaline phosphatase lab value.
\item[lab.value.ast:] AST lab value.
\item[lab.value.bilitot:] Total bilirubin lab value.
\item[lab.value.lipase:] Lipase lab value.
\item[lab.value.aptt:] APT hepatic enzime time lab value.
\item[lab.value.calcion:] Calcium lab value.
\item[lab.value.inr:] INR lab value.
\item[lab.value.lactate:] Lactate lab value.
\item[lab.value.magnes:] Magnesium lab value.
\item[lab.value.ph:] Plasma pH lab value
\item[lab.value.amylase:] Amylase lab value
\item[lab.value.bilidir:] Direct bilirubin lab value
\item[lab.value.phos:] Phosphourus lab value.
\item[lab.value.calcium:] Calcium lab value.
\item[lab.value.alt:] ALT hepatic enzime lab value.
\item[lab.value.ammonia:] Ammonia lab value.
\item[lab.value.pco2:] Arterial partial Co2 pressure lab value.
\item[lab.value.po2:] Arterial partial O2 pressure lab value.
\item[lab.value.albumin:] Albumin lab value.
\item[lab.value.sedrate:] Sedimentation rate lab value.
\item[lab.value.crp:] C reactive protein lab value.
\item[lab.value.egfr:] Estimated glomerular filtration rate lab value.
\item[charlson.score:] Charlson comorbidity count.
\item[cnt.hosp:] Hospitalizations count.
\item[ageyear:] Patient age.
\item[married:] Married = 1, Not Married = 0.
\item[male:] Male = 1, Female = 0.
\item[ethnicity:] Patient ethnicity taking on 1 of 10 possible categorical values
\item[vitals.map:] Mean arterial blood pressure
\item[vitals.si:] Shock index (heart rate divided by systolic arterial blood pressure)
\item[vitals.sbp:] Current (most recent in current hospitalization) systolic blood pressure.
\item[vitals.dbp:] Current (most recent in current hospitalization) diastolic blood pressure.
\item[vitals.hr:] Current (most recent in current hospitalization) heart rate.
\item[vitals.resp.rate:] Current (most recent in current hospitalization) respiratory rate.
\item[vitals.spo2:] Current (most recent in current hospitalization) oxygen saturation.
\item[vitals.temp:] Current (most recent in current hospitalization) body temperature.
\item[vitals.supp.oxygen:] Patient on supplemental oxygen = 1, otherwise = 0.
\item[kirkland.probability:] Current (most recent in current hospitalization) Kirkland probability index.
\item[episode.cnt:] Episode (uninterrupted stay in a general care bed)count
\item[los.episode:] Length of stay of current episode (i.e. time since current admission or transfer to current general care bed).
\item[vitals.weight:] Patient weight.
\item[dialysis.patient:] Patient on dialysis = 1, otherwise = 0.
\item[iv.sol.drug.dose.2hr:] Amount of fluids administered intravenously in the last 2 hours.
\item[iv.sol.drug.dose.4hr:] Amount of fluids administered intravenously in the last 2 hours.
\item[med.1hr.class8:] Class 8 medication administration in the last hour.
\item[med.2hr.class8:] Class 8 medication administration in the last 2 hours.
\item[med.4hr.class8:] Class 8 medication administration in the last 4 hours.
\item[med.duration.class1:] Class 1 medication administered recently and still considered active in the patient.
\item[med.duration.class2:] Class 2 medication administered recently and still considered active in the patient.
\item[med.duration.class3:] Class 3 medication administered recently and still considered active in the patient.
\item[med.duration.class4:] Class 4 medication administered recently and still considered active in the patient.
\item[med.duration.class5:] Class 5 medication administered recently and still considered active in the patient.
\item[med.duration.class6:] Class 6 medication administered recently and still considered active in the patient.
\item[med.duration.class7:] Class 7 medication administered recently and still considered active in the patient.
\item[med.duration.class8:] Class 8 medication administered recently and still considered active in the patient.
\item[med.duration.class9:] Class 9 medication administered recently and still considered active in the patient.
\item[med.duration.class10:] Class 10 medication administered recently and still considered active in the patient.
\item[med.duration.class11:] Class 11 medication administered recently and still considered active in the patient.
\item[med.duration.class12:] Class 12 medication administered recently and still considered active in the patient.
\item[med.duration.class13:] Class 13 medication administered recently and still considered active in the patient.
\item[med.duration.class14:] Class 14 medication administered recently and still considered active in the patient.
\item[med.duration.class15:] Class 15 medication administered recently and still considered active in the patient.
\item[med.duration.class16:] Class 16 medication administered recently and still considered active in the patient.
\item[med.duration.class17:] Class 17 medication administered recently and still considered active in the patient.
\item[med.duration.class18:] Class 18 medication administered recently and still considered active in the patient.
\item[med.duration.class19:] Class 19 medication administered recently and still considered active in the patient.
\item[med.duration.class20:] Class 20 medication administered recently and still considered active in the patient.
\item[med.duration.class21:] Class 21 medication administered recently and still considered active in the patient.
\item[med.duration.class22:] Class 22 medication administered recently and still considered active in the patient.
\item[med.duration.class23:] Class 23 medication administered recently and still considered active in the patient.
\item[med.duration.class24:] Class 24 medication administered recently and still considered active in the patient.
\item[med.duration.class25:] Class 25 medication administered recently and still considered active in the patient.
\item[med.duration.class26:] Class 26 medication administered recently and still considered active in the patient.
\item[med.duration.class27:] Class 27 medication administered recently and still considered active in the patient.
\item[med.duration.class28:] Class 28 medication administered recently and still considered active in the patient.
\item[med.duration.class29:] Class 29 medication administered recently and still considered active in the patient.
\item[med.duration.class30:] Class 30 medication administered recently and still considered active in the patient.
\item[med.duration.class31:] Class 31 medication administered recently and still considered active in the patient.
\item[med.duration.class32:] Class 32 medication administered recently and still considered active in the patient.
\item[med.duration.class33:] Class 33 medication administered recently and still considered active in the patient.
\item[med.duration.class34:] Class 34 medication administered recently and still considered active in the patient.
\item[med.duration.class35:] Class 35 medication administered recently and still considered active in the patient.
\item[med.duration.class36:] Class 36 medication administered recently and still considered active in the patient.
\item[med.duration.class37:] Class 37 medication administered recently and still considered active in the patient.
\item[med.duration.class38:] Class 38 medication administered recently and still considered active in the patient.
\item[med.duration.class39:] Class 39 medication administered recently and still considered active in the patient.
\item[med.duration.class40:] Class 40 medication administered recently and still considered active in the patient.
\item[med.duration.class41:] Class 41 medication administered recently and still considered active in the patient.
\item[med.duration.class42:] Class 42 medication administered recently and still considered active in the patient.
\item[med.duration.class43:] Class 43 medication administered recently and still considered active in the patient.
\item[med.duration.class44:] Class 44 medication administered recently and still considered active in the patient.
\item[med.duration.class45:] Class 45 medication administered recently and still considered active in the patient.
\item[med.duration.class46:] Class 46 medication administered recently and still considered active in the patient.
\item[med.duration.class47:] Class 47 medication administered recently and still considered active in the patient.
\item[med.duration.class48:] Class 48 medication administered recently and still considered active in the patient.
\item[med.duration.class49:] Class 49 medication administered recently and still considered active in the patient.
\item[med.duration.class50:] Class 50 medication administered recently and still considered active in the patient.
\item[med.duration.class51:] Class 51 medication administered recently and still considered active in the patient.
\item[med.duration.class52:] Class 52 medication administered recently and still considered active in the patient.
\item[med.duration.class53:] Class 53 medication administered recently and still considered active in the patient.
\item[med.duration.class54:] Class 54 medication administered recently and still considered active in the patient.
\item[med.duration.class55:] Class 55 medication administered recently and still considered active in the patient.
\item[med.duration.class56:] Class 56 medication administered recently and still considered active in the patient.
\item[med.duration.class57:] Class 57 medication administered recently and still considered active in the patient.
\item[med.duration.class58:] Class 58 medication administered recently and still considered active in the patient.
\item[med.duration.class59:] Class 59 medication administered recently and still considered active in the patient.
\item[med.duration.class60:] Class 60 medication administered recently and still considered active in the patient.
\item[med.duration.class61:] Class 61 medication administered recently and still considered active in the patient.
\item[med.duration.class62:] Class 62 medication administered recently and still considered active in the patient.
\item[med.duration.class63:] Class 63 medication administered recently and still considered active in the patient.
\item[med.duration.class64:] Class 64 medication administered recently and still considered active in the patient.
\item[med.avnode.duration:] Medications affecting the AV node.
\item[los.hours:] Current length of stay (time from current hospital admission to that point in time).
\item[med.tot.n:] Total number of medications currently prescribed to the patient.
\item[bmi:] Patient Body Mass Index.
\item[int.sbp.ivsol2hr:] vitals.sbp $\times$ iv.sol.drug.dose.2hr
\item[int.sbp.ivsol4hr:] vitals.sbp $\times$ iv.sol.drug.dose.4hr
\item[int.diuret.bun.creat:] Interaction term including lab.bun.to.creat and administration of diuretics.
\item[int.spo2.hemog:] vitals.spo2 $\times$ lab.value.hemog
\item[int.rr.spo2.o2flow:] vitals.resp.rate $\times$ vitals.spo2 $\times$ lab.value.o2flow
\item[int.neb.rr.spo2.opioids:] Interaction term including respiratory rate, spO2, and administration of nebulizers and opioids 
\item[int.hr.hemog.ratio:] vitals.hr $\slash$ lab.value.hemog
\item[prev.rrt.cnt:] Count of previous RRT calls.
\item[prev.code45.cnt:] Count of previous Codes (respiratory arrests).
\item[prev.xicu.cnt:] Count of previous ICU transfers.
\item[prev.outcome.cnt:] Count of previous RRT calls, Codes and ICU transfers
\item[lab.bun.to.creat:] lab.value.bun $\slash$ lab.value.creat
\item[meds.fluids:] Patient currently on intravenous fluids.
\item[int.rr.spo2.ratio:] vitals.rr $\slash$ vitals.spo2
\item[int.rr.spo2.opioids:] vitals.rr $\times$ vitals.spo2 $\times$ use of opioids
\item[general:] General care patient or patient on telemetry.
\item[max24.vitals.sbp:] Maximum value of vitals.sbp in the past 24 hours.
\item[min24.vitals.sbp:] Minimum value of vitals.sbp in the past 24 hours.
\item[max24.vitals.dbp:] Maximum value of vitals.dbp in the past 24 hours.
\item[min24.vitals.dbp:] Minimum value of vitals.dbp in the past 24 hours.
\item[max24.spb.dbp:] Maximum value of vitals.sbp $\times$ vitals.dbp in the past 24 hours.
\item[min24.spb.dbp:] Minimum value of vitals.sbp $\times$ vitals.dbp in the past 24 hours.
\item[max24.vitals.resp.rate:] Maximum value of vitals.resp.rate in the past 24 hours.
\item[min24.vitals.resp.rate:] Minimum value of vitals.resp.rate in the past 24 hours.
\item[max24.vitals.hr:] Maximum value of vitals.hr in the past 24 hours.
\item[min24.vitals.hr:] Minimum value of vitals.hr in the past 24 hours.
\item[max24.vitals.spo2:] Maximum value of vitals.spo2 in the past 24 hours.
\item[min24.vitals.spo2:] Minimum value of vitals.spo2 in the past 24 hours.
\item[max24.int.hr.hemog.ratio:] Maximum value of int.hr.hemog.ratio in the past 24 hours.
\item[min24.int.hr.hemog.ratio:] Minimum value of int.hr.hemog.ratio in the past 24 hours.
\item[max24.int.rr.spo2.ratio:] Maximum value of int.rr.spo2.ratio in the past 24 hours.
\item[min24.int.rr.spo2.ratio:] Minimum value of int.rr.spo2.ratio in the past 24 hours.
\item[range24.vitals.sbp:] max24.vitals.sbp $-$ min24.vitals.sbp
\item[range24.vitals.dbp:] max24.vitals.dbp $-$ min24.vitals.dbp
\item[range24.vitals.resp.rate:] max24.vitals.resp.rate $-$ min24.vitals.resp.rate
\item[range24.vitals.hr:] max24.vitals.hr $-$ min24.vitals.hr
\item[range24.vitals.spo2:] max24.vitals.spo2 $-$ min24.vitals.spo2
\item[range24.int.hr.hemog.ratio:] max24.int.hr.hemog.ratio $-$ min24.int.hr.hemog.ratio
\end{description}

\end{appendix}

\end{document}